\documentclass[aps,twocolumn]{revtex4}
\usepackage{epsfig,graphics,amssymb,amsmath,subeqnarray}
\usepackage{color}

\def \H {\mathbf{H}}
\def \f {\mathbf{f}}
\def \t {\mathbf{t}}
\def \u {\mathbf{u}}
\def \U {\mathbf{U}}
\def \r {\mathbf{r}}
\def \f {\mathbf{f}}
\def \Sp {\text{Sp}}

\def \vb {\mathbf{v}_b}
\def \h {\text{h}}
\def \H {\mathbf{H}}

\def \M {\mathbf{M}}
\def \N {\mathbf{N}}

\def \OOmega {\boldsymbol\omega}

\begin{document}

\title{High-Speed Propulsion of Flexible Nanowire Motors: Theory and Experiments}

\author{On Shun Pak\footnote[1]{These authors contributed equally to this work.}}
\affiliation{
Department of Mechanical and Aerospace Engineering, 
University of California San Diego,
9500 Gilman Drive, La Jolla CA 92093-0411, USA.}
\author{Wei Gao\footnotemark[1]}
\affiliation{
Department of Nanoengineering, University of California San Diego, 9500 Gilman Drive, La Jolla CA 92093-0448, USA.}
\author{Joseph Wang \footnote[3]{Email: josephwang@ucsd.edu}}
\affiliation{
Department of Nanoengineering, University of California San Diego, 9500 Gilman Drive, La Jolla CA 92093-0448, USA.}
\author{Eric Lauga \footnote[2]{Email: elauga@ucsd.edu}}
\affiliation{
Department of Mechanical and Aerospace Engineering, 
University of California San Diego,
9500 Gilman Drive, La Jolla CA 92093-0411, USA.}

\date{\today}

\begin{abstract}

Micro/nano-scale propulsion has attracted considerable recent attention due to its promise for biomedical applications such as targeted drug delivery. In this paper, we report on a new experimental design and theoretical modelling of high-speed fuel-free magnetically-driven propellers which exploit the flexibility of nanowires for propulsion. These readily prepared nanomotors display both high dimensional propulsion velocities (up to $\approx 21 \mu$m/s) and dimensionless speeds (in body lengths per revolution) when compared with natural microorganisms and other artificial propellers. Their propulsion characteristics  are studied theoretically using an elastohydrodynamic model which takes into account the elasticity of the nanowire and its hydrodynamic interaction with the fluid medium. The critical role of flexibility in this mode of propulsion is illustrated by simple physical arguments, and is quantitatively investigated with the help of an asymptotic analysis for small-amplitude swimming. The theoretical predictions are then compared with experimental measurements and we obtain good agreement. Finally, we demonstrate the operation of these nanomotors in a real biological environment (human serum), emphasizing the robustness of their propulsion performance and their promise for biomedical applications.

\end{abstract}

\maketitle

\section{Introduction}

Micro/nano-scale propulsion in fluids is challenging due to the absence of the inertial forces exploited by  biological organisms on macroscopic scales. The difficulties are summarized by Purcell's famous ``scallop theorem" \cite{purcell77}, which states that a reciprocal motion (a deformation with time-reversal symmetry)  cannot lead to any net propulsion at low Reynolds numbers. The Reynolds number, $\text{Re} = \rho U L/\mu$, measures the relative importance of inertial to viscous forces, where $\rho$ and $\mu$ are the density and shear viscosity of the fluid, while $U$ and $L$ are the characteristic velocity and length scales of the self-propelling body. Natural microorganisms inhabit a world where $\text{Re} \sim 10^{-5}$ (flagellated bacteria) to $10^{-2}$ (spermatozoa) \cite{brennen,childress}, and they achieve their propulsion  by propagating traveling waves along their flagella (or rotating them) to break the time-reversibility requirement, and hence escape the constraints of the scallop theorem \cite{brennen, lauga2}. Because of the potential of nano-sized machines in future biomedical applications \cite{nelson10}, such as targeted drug delivery and microsurgery, interdisciplinary efforts by scientists and engineers have recently resulted in major advances in the design and fabrication of artificial micro/nano-scale locomotive systems \cite{wang09, mirkovic, mallouk,ebbens}.

\begin{table*}
\small
  \caption{\label{table:swimmers} Comparison between natural micro-organisms and different experimentally realized externally-powered fuel-free propellers, divided into three main categories: (a) flexible propellers;  (b) rigid helical propellers; (c) surface walkers (i.e.~requiring a surface for propulsion). We report the maximum dimensional speeds $U_{\text{max}}$ and the maximum dimensionless speeds $\tilde{U}_{\text{max}}=U/Lf$, and their corresponding characteristic lengths $L$ and actuation frequencies $f$. ($^*$:  estimated from the dimensionless speed in Ref.~\cite{dreyfus}).}
  \label{tbl:example}
  \begin{tabular*}{\textwidth}{@{\extracolsep{\fill}}llllll} \hline \hline

\multicolumn{1}{l}{} & 
\multicolumn{1}{c}{} &  
\multicolumn{2}{c}{Maximum dimensional speed} &
\multicolumn{2}{c}{Maximum dimensionless speed}\\

\multicolumn{1}{l}{Type of Propellers} & 
\multicolumn{1}{c}{Schematic/Micrograph} &  
\multicolumn{2}{c}{$U_{\text{max}} \ [\mu$m/s]} &
\multicolumn{2}{c}{$\tilde{U}_{\text{max}}=U/Lf \ (\times 10^{-3})$} \\

    \hline

\multicolumn{1}{l}{\raisebox{6ex}{\textit{Escherichia coli} \cite{turner}}} &
\multicolumn{1}{c}{{\includegraphics[width=0.1\textwidth]{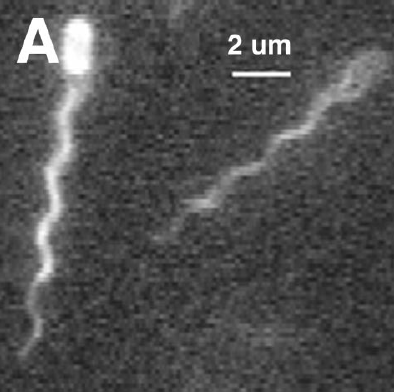}}}&
\multicolumn{1}{l}{\raisebox{6ex}{$U \approx 30 \mu$m/s}}&
\multicolumn{1}{l}{\raisebox{6ex}{\parbox{2.1cm}{\begin{align}L &= 10 \mu \text{m} \notag \\ f &= 100\text{Hz}  \notag\\ \tilde{U} &\approx 30\notag  \end{align}}}}&
\multicolumn{1}{l}{}&
\multicolumn{1}{l}{}\\

\hline

\multicolumn{1}{l}{\raisebox{5ex}{Flexible propeller \cite{dreyfus}}} &
\multicolumn{1}{c}{\raisebox{3ex}{\includegraphics[width=0.15\textwidth]{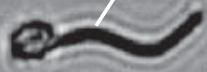}}}&
\multicolumn{1}{l}{}&
\multicolumn{1}{l}{}&
\multicolumn{1}{l}{\raisebox{5ex}{$\tilde{U}_{\text{max}} \approx 94$}}&
\multicolumn{1}{l}{\raisebox{5ex}{\parbox{2.1cm}{\begin{align}L &= 24 \mu \text{m} \notag \\ f &= 10\text{Hz}  \notag\\ U &\approx 22 \mu \text{m/s}^* \notag  \end{align}}}}\\

\multicolumn{1}{l}{\raisebox{5ex}{Flexible propeller \cite{gao}}}&
\multicolumn{1}{c}{\includegraphics[width=0.15\textwidth]{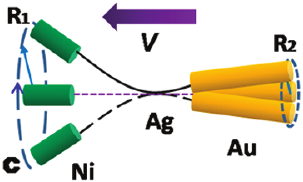}}&
\multicolumn{1}{l}{\raisebox{5ex}{$U_{\text{max}}= 6 \mu$m/s}}&
\multicolumn{1}{l}{\raisebox{5ex}{\parbox{2.1cm}{\begin{align}L &= 6.5 \mu \text{m} \notag \\ f &= 15\text{Hz}  \notag\\ \tilde{U} &= 62 \notag  \end{align}}}}&
\multicolumn{1}{l}{\raisebox{5ex}{$\tilde{U}_{\text{max}} = 77$}}&
\multicolumn{1}{l}{\raisebox{5ex}{\parbox{2.1cm}{\begin{align}L &= 6.5 \mu \text{m} \notag \\ f &= 7\text{Hz}  \notag\\ U &= 3.5 \mu \text{m/s}\notag  \end{align}}}}\\

\multicolumn{1}{l}{\raisebox{8ex}{\parbox{3cm}{Flexible propeller \\ (the current paper)}}} &
\multicolumn{1}{c}{\includegraphics[width=0.15\textwidth]{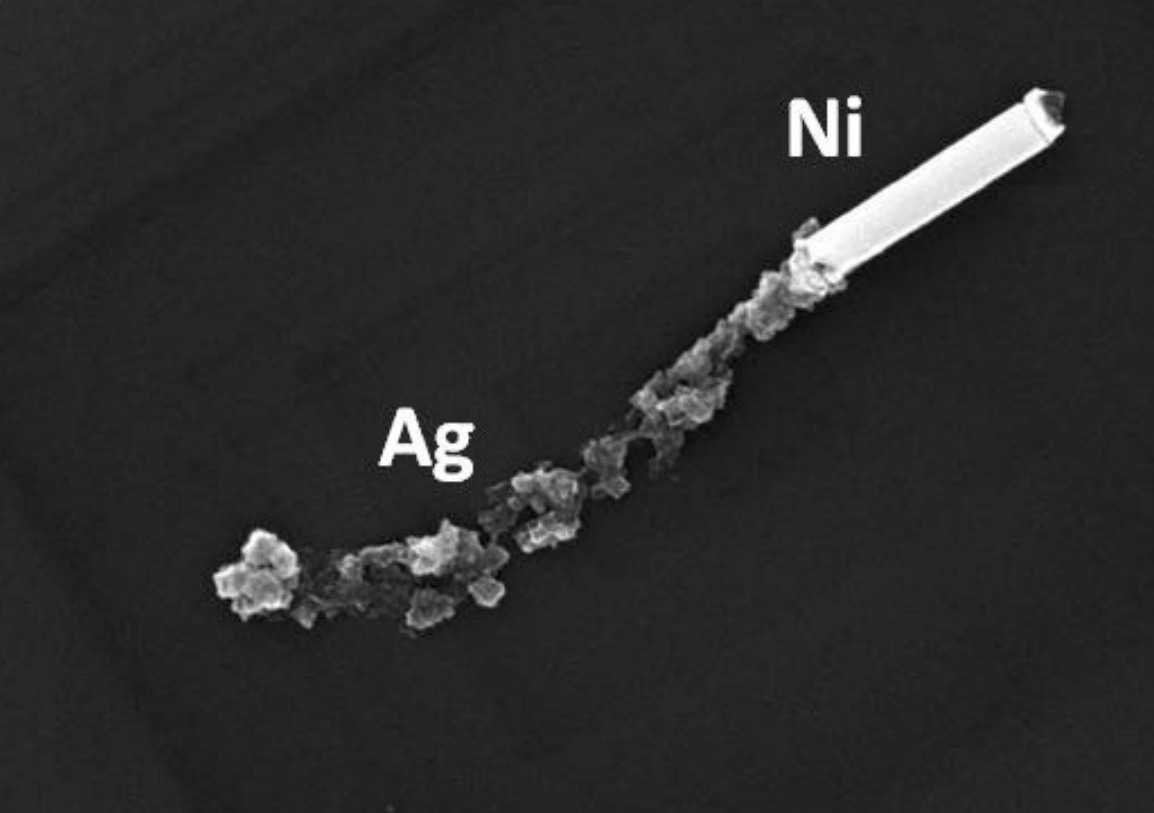}}&
\multicolumn{1}{l}{\raisebox{8ex}{$U_{\text{max}}= 21 \mu$m/s}}&
\multicolumn{1}{l}{\raisebox{8ex}{\parbox{2.1cm}{\begin{align}L &= 5.8 \mu \text{m} \notag \\ f &= 35\text{Hz}  \notag\\ \tilde{U} &= 103 \notag  \end{align}}}}&
\multicolumn{1}{l}{\raisebox{8ex}{$\tilde{U}_{\text{max}} = 164$}}&
\multicolumn{1}{l}{\raisebox{8ex}{\parbox{2.1cm}{\begin{align}L &= 5.8 \mu \text{m} \notag \\ f &= 15\text{Hz}  \notag\\ U &= 14.3 \mu \text{m/s}\notag  \end{align}}}}\\

\hline

\multicolumn{1}{l}{\raisebox{4ex}{Helical propeller \cite{zhang09b}}}&
\multicolumn{1}{c}{\includegraphics[width=0.15\textwidth]{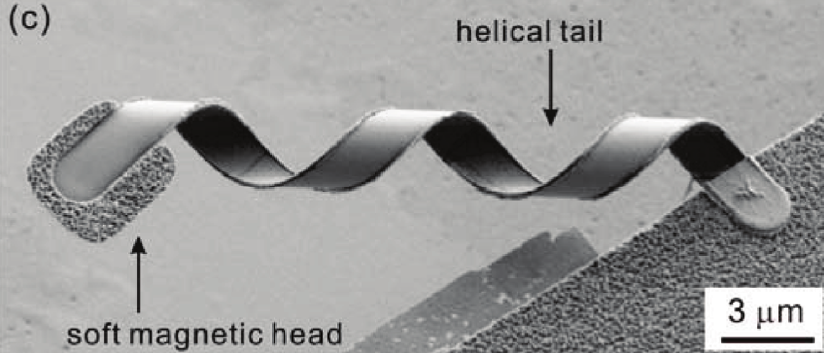}}&
\multicolumn{1}{l}{\raisebox{4ex}{$U_{\text{max}}= 18 \mu$m/s}}&
\multicolumn{1}{l}{\raisebox{4ex}{\parbox{2.1cm}{\begin{align}L &= 38 \mu \text{m} \notag \\ f &= 30\text{Hz}  \notag\\ \tilde{U} &= 16 \notag  \end{align}}}}&
\multicolumn{1}{l}{\raisebox{4ex}{$\tilde{U}_{\text{max}} = 21$}}&
\multicolumn{1}{l}{\raisebox{4ex}{\parbox{2.1cm}{\begin{align}L &= 38 \mu \text{m} \notag \\ f &= 10\text{Hz}  \notag\\ U &= 8 \mu \text{m/s}\notag  \end{align}}}}\\

\multicolumn{1}{l}{\raisebox{7ex}{Helical propeller \cite{ghosh}}}&
\multicolumn{1}{c}{\includegraphics[width=0.1\textwidth]{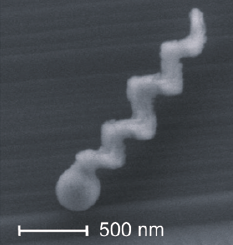}}&
\multicolumn{1}{l}{\raisebox{7ex}{$U_{\text{max}}= 40 \mu$m/s}}&
\multicolumn{1}{l}{\raisebox{7ex}{\parbox{2.1cm}{\begin{align}L &= 2 \mu \text{m} \notag \\ f &= 150\text{Hz}  \notag\\ \tilde{U} &= 133 \notag  \end{align}}}}&
\multicolumn{1}{l}{}&
\multicolumn{1}{l}{}\\

\hline

\multicolumn{1}{l}{\raisebox{6ex}{Surface walker \cite{tierno08,tierno10}}}&
\multicolumn{1}{c}{\raisebox{-1ex}{\includegraphics[width=0.1\textwidth]{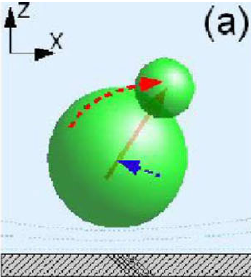}}}&
\multicolumn{1}{l}{\raisebox{6ex}{$U_{\text{max}}= 3.5 \mu$m/s}}&
\multicolumn{1}{l}{\raisebox{6ex}{\parbox{2.1cm}{\begin{align}L &= 4 \mu \text{m} \notag \\ f &= 15\text{Hz}  \notag\\ \tilde{U} &= 58 \notag  \end{align}}}}&
\multicolumn{1}{l}{\raisebox{6ex}{$\tilde{U}_{\text{max}} = 80$}}&
\multicolumn{1}{l}{\raisebox{6ex}{\parbox{2.1cm}{\begin{align}L &= 4 \mu \text{m} \notag \\ f &= 10\text{Hz}  \notag\\ U &= 3.2 \mu \text{m/s}\notag  \end{align}}}}\\

\multicolumn{1}{l}{\raisebox{5ex}{Surface walker \cite{sing}}}&
\multicolumn{1}{c}{\raisebox{-1ex}{\includegraphics[width=0.12\textwidth]{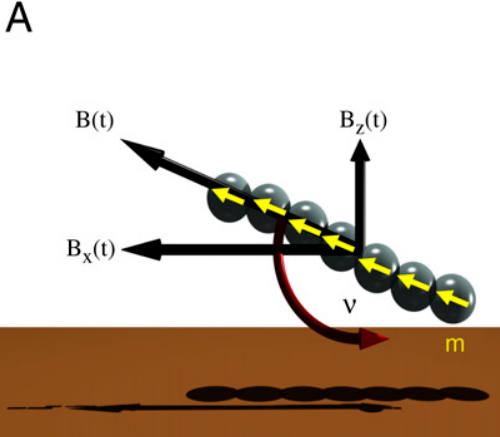}}}&
\multicolumn{1}{l}{\raisebox{5ex}{$U_{\text{max}}= 12 \mu$m/s}}&
\multicolumn{1}{l}{\raisebox{5ex}{\parbox{2.1cm}{\begin{align}L &= 3 \mu \text{m} \notag \\ f &= 32\text{Hz}  \notag\\ \tilde{U} &= 125 \notag  \end{align}}}}&
\multicolumn{1}{l}{}&
\multicolumn{1}{l}{}\\

\multicolumn{1}{l}{\raisebox{8ex}{Surface walker \cite{zhang10}}}&
\multicolumn{1}{c}{\raisebox{1ex}{\includegraphics[width=0.12\textwidth]{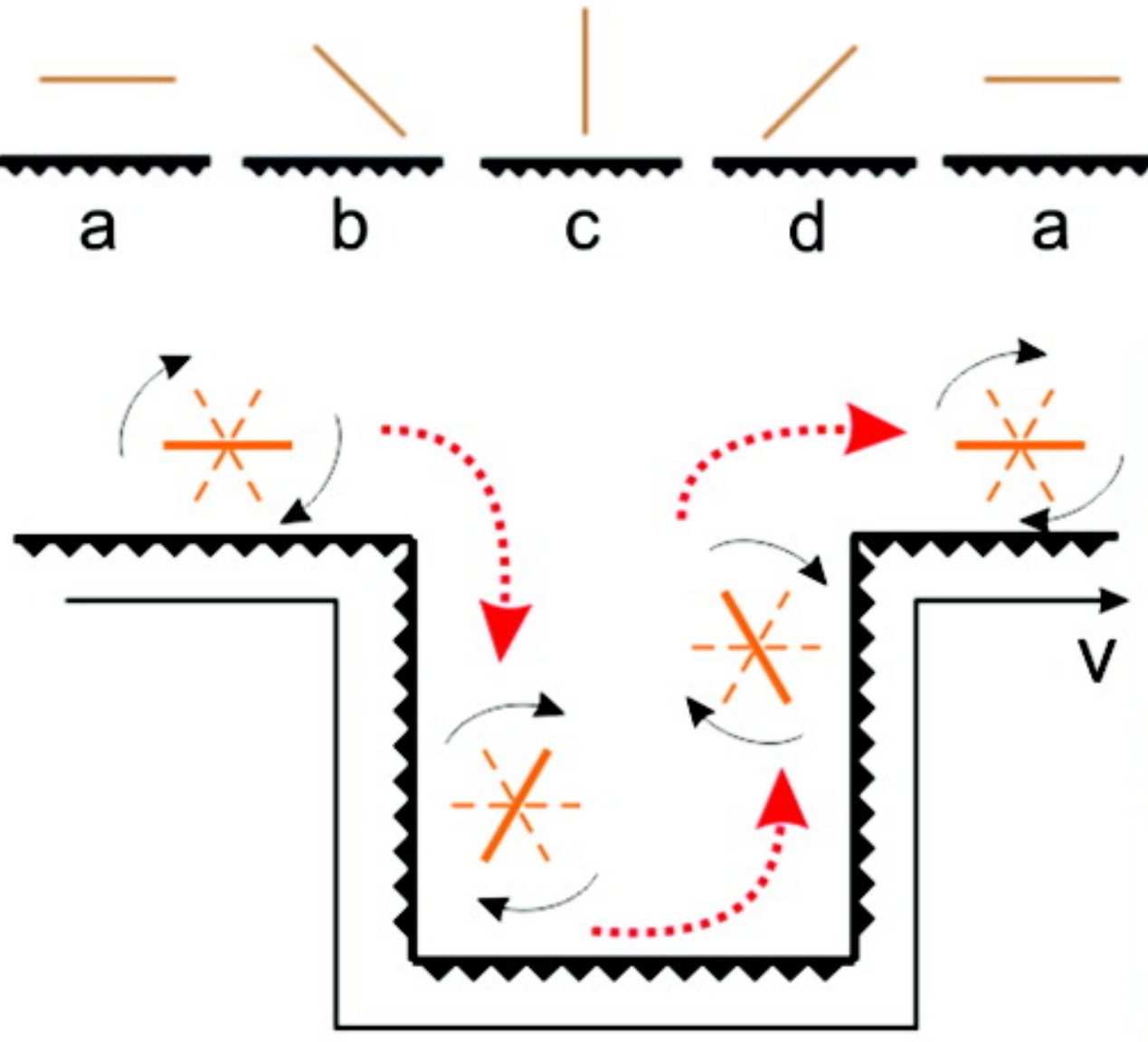}}}&
\multicolumn{1}{l}{\raisebox{8ex}{$U_{\text{max}}= 37 \mu$m/s}}&
\multicolumn{1}{l}{\raisebox{8ex}{\parbox{2.1cm}{\begin{align}L &= 12 \mu \text{m} \notag \\ f &= 35\text{Hz}  \notag\\ \tilde{U} &= 88 \notag  \end{align}}}}&
\multicolumn{1}{l}{\raisebox{8ex}{$\tilde{U}_{\text{max}} = 90$}}&
\multicolumn{1}{l}{\raisebox{8ex}{\parbox{2.1cm}{\begin{align}L &= 4 \mu \text{m} \notag \\ f &= 47\text{Hz}  \notag\\ U &= 17 \mu \text{m/s}\notag  \end{align}}}}\\
  \end{tabular*}
\end{table*}

Broadly speaking, these micro/nano-propellers can be classified into two categories, namely chemically-powered nanomotors \cite{wang09, mirkovic, mallouk} and externally-powered propellers \cite{ebbens}. Chemically-powered nanomotors generally deliver higher propulsion speeds, but due to the requirements for chemical fuels and reactions, their applications in real biological environments  face a number of challenges. Externally-powered propellers are often actuated by external magnetic fields. Note that these externally-powered locomotive systems are often referred to micro- or nano-swimmers in the literatures, but strictly speaking, they do not represent true self-propulsion because of the presence of non-zero external torques. In this paper, we reserve the terminology, ``swimmers", to force-free and torque-free self-propelling bodies and refer to externally-powered locomotive systems as propellers, or motors.

According to their propulsion mechanisms, externally powered propellers can be further categorized into three groups (see the summary presented in Table~\ref{table:swimmers}). The first group includes helical propellers \cite{zhang09b, ghosh}, as inspired by helical bacterial flagella \cite{turner}, which propel upon rotation imposed by external magnetic fields. The second group of propellers relies on a surface to break the spatial symmetry and provide one additional degree of freedom to escape the constraints from the scallop theorem, and hence are termed surface walkers \cite{tierno08,tierno10, sing, zhang10}. Finally, the third type of propellers, referred to as flexible propellers, exploits the deformation of flexible filaments for propulsion. The new nanomotor presented in this paper falls into this category. Dreyfus \textit{et al.} \cite{dreyfus} were the first to realize the idea experimentally by fabricating a $24 \mu$m long propeller based on a flexible filament, made of paramagnetic beads linked by DNA, and attached to a red blood cell. Actuation was distributed along the filament by the paramagnetic beads; the presence of the red blood cell broke the front-back symmetry, and allowed the propagation of a traveling wave along the filament, leading to propulsion. Recently, Gao \textit{et al.} \cite{gao} proposed a  flexible nanowire motor made of only metallic nanowires (with three segments of Au, Ag and Ni) readily prepared using a template electrodeposition approach,  and able to swim at speeds of up to $U \approx 6\mu$m/s for a size of $L \approx 6.5 \mu$m.  In contrast to the propeller proposed by Dreyfus \textit{et al.} \cite{dreyfus}, the actuation in the device of Gao \textit{et al.} \cite{gao} acted only on the magnetic Ni portion of the filament (the head), while the rest of nanomotor was passive.

In the current paper, we present both a new design and a theoretical modelling approach for a flexible nanowire motor which offers an improved propulsion performance (up to $U \approx 21\mu$m/s at an actuation frequency ($f$) of $35$Hz), approaching thus the speed of natural microscopic swimmers, such as \textit{Escherichia coli} ($U \approx 30 \mu$m/s at $f=100$Hz) \cite{turner} while using a lower frequency. The effect of size and frequency can be scaled off by nondimensionalizing the propulsion speed by the intrinsic velocity scale (the product of body length and frequency, $Lf$) to obtain a dimensionless propulsion speed, $U/Lf$, which can be interpreted as the number of body lengths travelled per revolution of actuation (or also referred to as the stride length in terms of body length in the biomechanics literature). The  nanomotor put forward in this paper displays remarkable dimensionless propulsion speeds compared with natural microorganisms and other artificial locomotive systems (see summary of the literature and the current results in Table~\ref{table:swimmers}).

After presenting the experimental method and its performance, we study the propulsion characteristics of this new high-speed flexible nanomotor theoretically via an analytical model. The critical role of flexibility in this mode of propulsion is established first using simple physical arguments, followed by an asymptotic analysis which predicts the filament shape and propulsion speed in different physical regimes. The theoretical predictions are compared with experimental measurements and we obtain good agreement. The improved propulsion performance of the new fuel-free nanowire motor makes it attractive for future biomedical applications, which we further illustrate by demonstrating the performance of the propulsion mechanism in an untreated human serum sample.

\section{High-speed propulsion}

\subsection{Nanomotor design and fabrication}
\begin{figure*}
\centering
\includegraphics[width=0.87\textwidth]{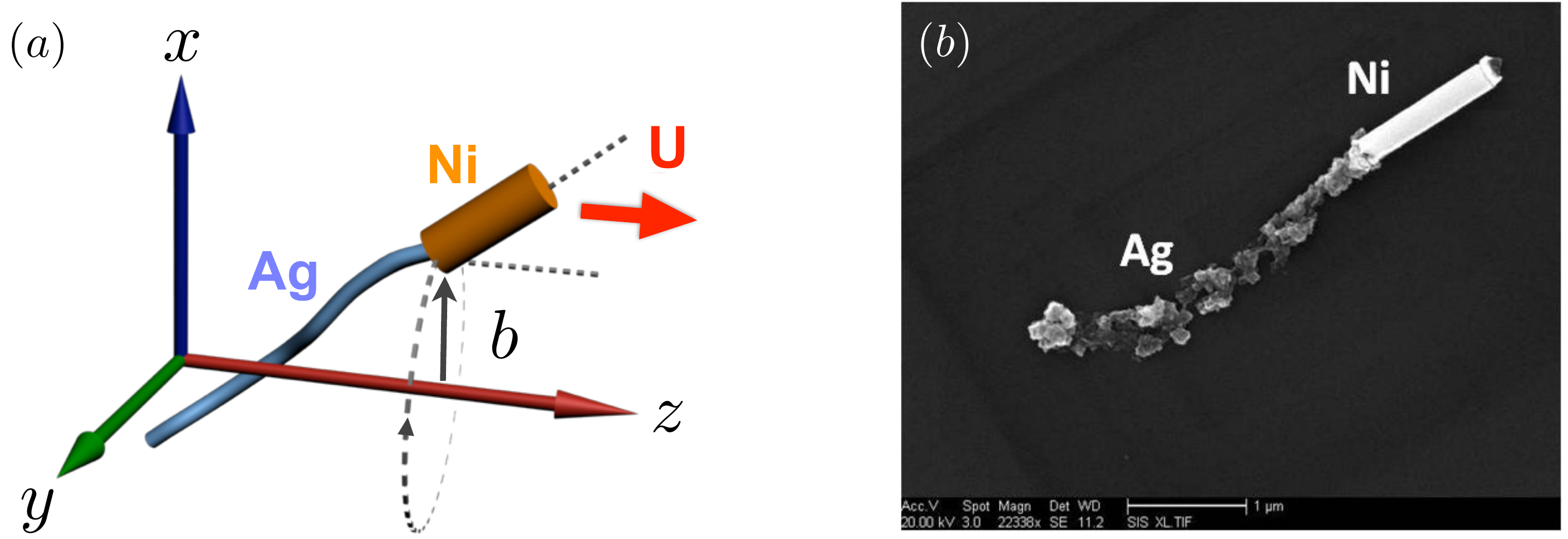}
\caption{\label{fig:setup}(a) Schematic representation of a Ni-Ag nanowire motor, and notation for the model. (b) Scanning Electron Microscopy (SEM) image showing the topography of Ni-Ag nanowire which was partially dissolved in 5\% H$_2$O$_2$ for 1 minute.}
\end{figure*}

The nanowire motors described in this paper were prepared using a common template-directed electrodeposition protocol. In contrast to the previous three-segment (Ni-Ag-Au) design 
by Gao \textit{et al.} \cite{gao}, the new design relies primarily on a 1.5$\mu$m-long Ni head and a 4 $\mu$m-long flexible Ag tail (see a Scanning Electron Microscopy (SEM) image in  Fig.~\ref{fig:setup}b). A 0.3$\mu$m-long Au segment was also included (adjacent to the Ni segment) to protect the Ni segment from acid etching during the dissolution of the Cu sacrificial layer, and to allow functionalizing the motor with different types of biomolecules and cargos. Both the Ni and Au segments have a diameter of 200nm. While the Ni segment has a length of 1.5$\mu$m useful to generate sufficient magnetic torques, only a very short segment of Au (0.3$\mu$m) was used to minimize the overall fluid drag of the nanomotor. Flexibility of the silver segment (Fig.~\ref{fig:setup}b) was achieved by its partial dissolution in hydrogen peroxide solution  \cite{gao}. The dissolution step leads also to hydroxyl products that chemisorb on the Ag surface and result in AgOH and Ag$_2$O surface products. The dissolved Ag filament had a reduced diameter of approximately 100nm. For the hydrodynamic model considered in this paper, the rigid short Au segment is hydrodynamically indistinguishable from the rigid Ni segment, and hence the Ni and Au segments are considered in the model as a single rigid 1.8$\mu$m-long segment (1.5$\mu$m Ni+  0.3$\mu$m Au), i.e.~the nanomotor has a total length of 5.8$\mu$m.

The speed of a nanomotor was measured using MetaMorph 7.6 software (Molecular Devices, Sunnyvale, CA), capturing movies at a frame rate of 30 frames per sec. The trajectory  was tracked using a Metamorph tracking module and the results were statistically analyzed using Origin software. The speed measured in this manner is a time-averaged distance travelled per unit time. The measurements were performed when the nanomotors had reached an equilibrium position (in which case the image of the nanowire would stay focused under the microscope), which leads therefore to the time-averaged measurement of $U$ in the laboratory frame. The equilibrium distance between the nanomotor and the bottom surface was estimated, by varying the focal plane of the microscope, to be at the scale of a few microns. The detailed experimental procedures can be found in ESI\dag.

\subsection{Propulsion performance}

\begin{figure}[t]
\centering
\includegraphics[width=0.27\textwidth]{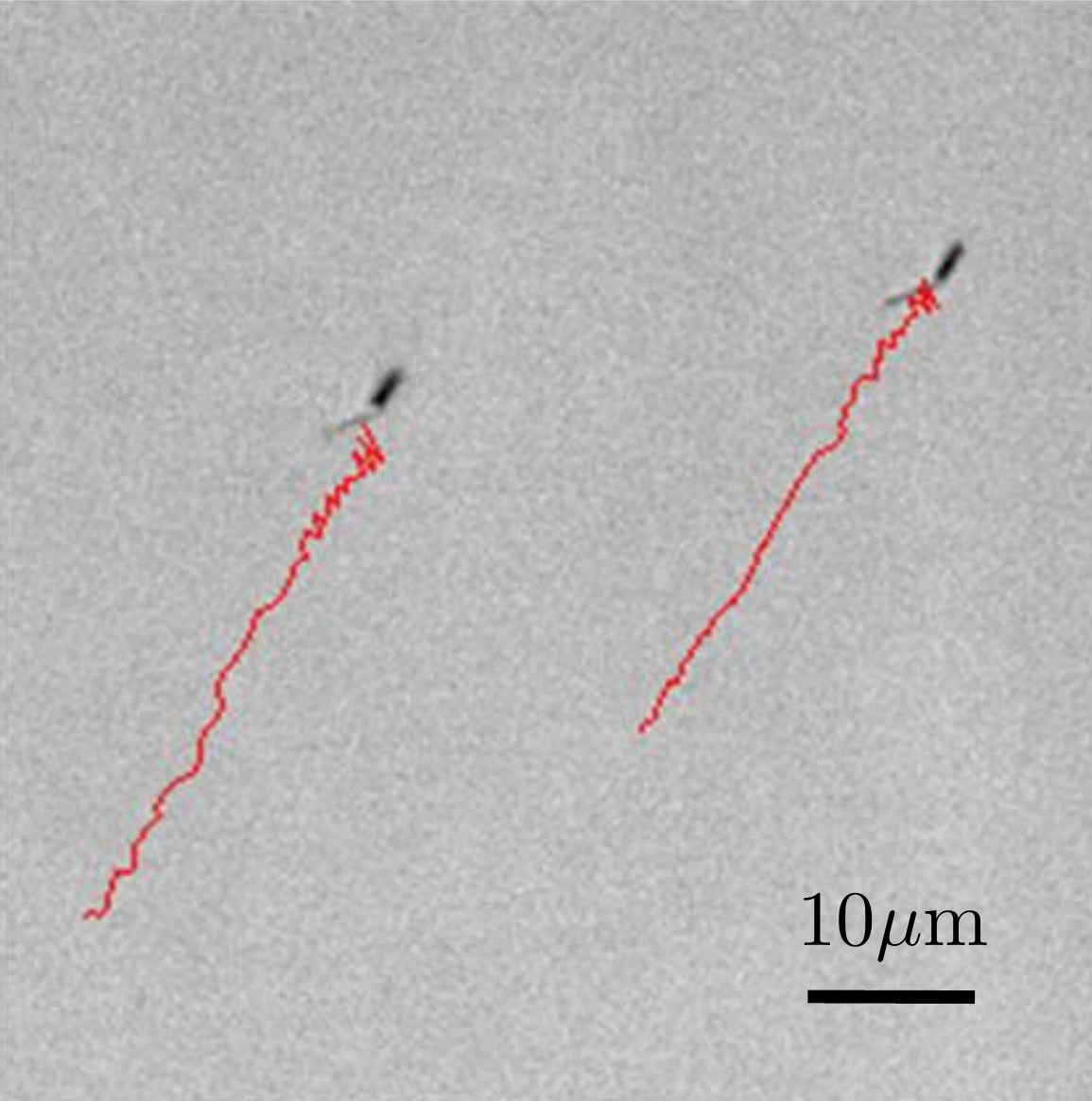}
\caption{\label{fig:comoving} Two identical nanomotors swimming under the same magnetic field at a frequency $f=20$Hz. The red lines display the  superimposed location of the nanomotors over a  2-second interval.}
\end{figure}

The flexible nanomotors were driven by a magnetic field with an unsteady component of amplitude $H_1$, rotating sinusoidally in a plane perpendicular to a constant component, $H_0$. The magnetic field precessed about the direction of the constant magnetic field at an angular frequency $\Omega=2\pi f$. The nanomotor was observed to propel unidirectionally (straight trajectories) in the direction of the constant magnetic field. In Fig.~\ref{fig:comoving} we show two nearby identical  nanomotors under the actuation of the external magnetic field at $f = 20$Hz (see Video 1\dag). These two nanowires propel  at essentially the same swimming speed along the same direction (the red lines are their trajectories in a period of 2 seconds), illustrating the stability of this mode of propulsion. For helical propellers \cite{zhang09b, ghosh}, swimming is due to the rotation of rigid chiral objects and hence the swimming kinematics scales linearly with the applied field: a reversal of the direction of rotation of the magnetic field leads to propulsion in the opposite direction for these rigid chiral objects. In contrast, the flexible nanowire motors here exhibit uni-directional swimming, independent of the rotational direction of the external magnetic field. This is due to the nonlinear swimming kinematics arising from the flexibility of the nanowire. This simple test illustrates the fundamental difference between the propulsion of rigid chiral objects and flexible propellers. In our case, the direction of swimming can be controlled by altering the orientation of the axial constant component of the magnetic field, $H_0$. 

We further show in Fig.~\ref{fig:TE}(a) the trajectories of the same nanomotor at different frequencies (see captions for details) over a 3-second period (see Video 2\dag). Upon the settings $H_1 = 10$G, $H_0 = 9.5$G, and $f = 15$Hz, we are able to achieve a propulsion speed of $U =14.3 \pm 2.46 \mu$m/s. The speed of 20 different nanomotors were measured, with all other experimental conditions kept fixed; the values of the swimming speeds,  $U$, reported in this paper are averaged quantities over these different nanomotors. One meaningful method of comparing the propulsion speed between various propeller designs consists in  scaling the speed with the only intrinsic characteristic velocity scale of the propeller $L f$, where $L$ is a characteristic body length, and $f$ is a characteristic frequency. This allows to quantify the distance travelled by the propeller in terms of body lengths per revolution of rotation. \textit{Escherichia coli} bacteria \cite{turner} typically propel with $U/ Lf \approx 0.03$ body lengths per revolution, while the flexible nanomotor reported here was able to travel $0.164$ body lengths per revolution at $f=15$Hz (see Table~\ref{table:swimmers} for detailed comparison).  The maximum dimensional speed achieved was $U =20.8 \pm 3.07 \mu$m/s with $f=35$Hz, corresponding in that case to $\approx 0.1$ body lengths per revolution. We   then experimentally measured the speed-frequency characteristics of these flexible nanowire motors (results shown as symbols in Fig.~\ref{fig:TE}b). In the next section we present a  simple physical model  for the locomotion of flexible nanomotors, and compare our theoretical predictions with these experimental measurements.

\section{A minimal model for flexible nanomotors}
\subsection{Chiral propulsion}
In this section, we illustrate the working principles of the flexible nanowire motors. First, we establish that it is essential for the nanowire to deform in a {chiral} fashion in order to achieve propulsion. 

\begin{figure}
\centering
\includegraphics[width=0.45\textwidth]{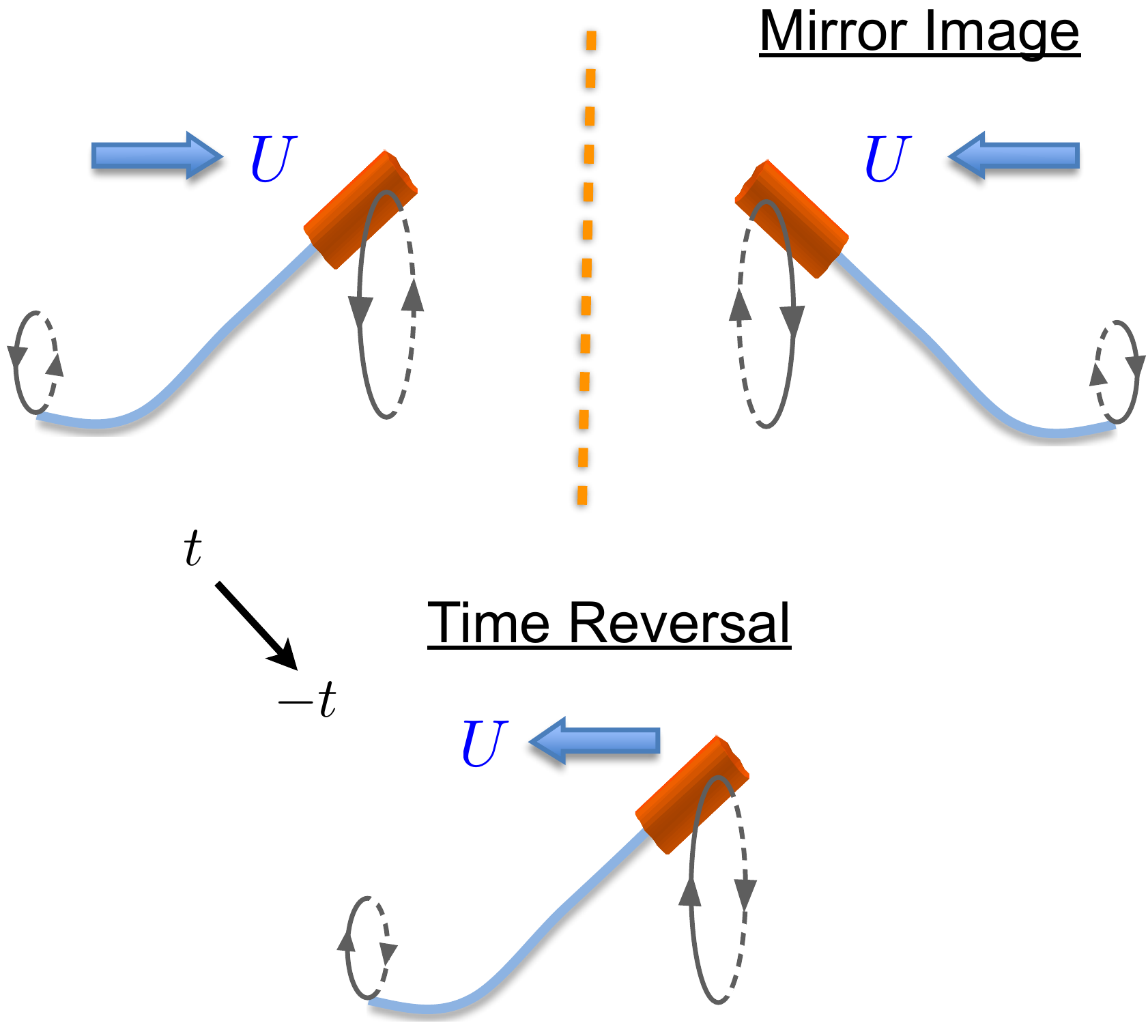}
\caption{\label{fig:symmetry}Physical explanation of the necessity of chiral deformation in achieving propulsion.
If the deformation is not chiral, the  kinematics of the mirror image of the nanowire is identical to the time-reversed kinematics, leading to $\mathbf{U}=\mathbf{0}$.}
\end{figure}

For low Reynolds number incompressible flows, the governing equations are the Stokes equation $\nabla p = \mu \nabla^2 \mathbf{u}$, and the continuity equation $\nabla \cdot \mathbf{u}=0$, where $\mu$ is the shear viscosity, and $p$ and $\mathbf{u}$ are the fluid pressure and the velocity field respectively. Two properties of the Stokes equation can be used to deduce the necessity of the nanowire being chiral in order to achieve net swimming, as shown by Childress \cite{childress}. First, it can be shown that the mirror image of a Stokes flow is also a Stokes flow. Therefore, suppose a nanowire swims with a velocity $\mathbf{U}$ along its rotation axis, then its mirror image will also swim at the same velocity $\mathbf{U}$ (see Fig.~\ref{fig:symmetry}). Second, since time does not appear in Stokes equation, it only enters the problem as a parameter through the boundary conditions. This leads to the time reversibility of the Stokes equation, meaning that the velocity field $\mathbf{u}$ reverses its sign upon a $t \rightarrow -t$ time reversal. In the context of our nanowire motors, suppose the nanowire propels at a velocity $\mathbf{U}$, then when time is reversed, the nanowire will propel at a velocity $-\mathbf{U}$ (Fig.~\ref{fig:symmetry}). If the deformation of the nanowire is not chiral, the mirror image of the nanowire can be superimposed with the original nanowire, and the only thing reversed in the mirror image is the rotational kinematics (\textit{i.e.}~if the original nanowire rotates clock-wisely, its mirror image will have exactly the same shape but rotates counter-wisely; note that the translational velocity is unchanged in the mirror image). In this case, one can also notice that the kinematics in the mirror image is the same as a time reversal of the original kinematics, except that the translational velocity is also reversed for the case of time-reversal ($-\mathbf{U}$, due to the time-reversibility of Stokes flows). In other words, we have now two nanowires (a mirror-imaged nanowire and a time-reversed nanowire) having exactly the same deformation kinematics but with opposite  translational velocity ($-\mathbf{U}=\mathbf{U}$), and therefore we conclude that this can happen only if the translational velocity is identically zero ($\mathbf{U}=\mathbf{0}$). Therefore non-chiral deformation cannot lead to net propulsion. This simple physical argument shows that a combination of rotational actuation and nanowire flexibility is critical for this mode of propulsion. Recently, the dynamics of tethered elastic filaments  actuated by  precessing magnetic fields has been studied \cite{kim,manghi,wada, coq, coq09,qian,coq10,downton} and chiral deformation along the filament has been found to produce propulsive force and fluid pumping. The swimming behaviours of an untethered flexible magnetic filament displaying chiral deformation was also addressed computationally \cite{keaveny}.

\subsection{Model setup}

Next we show that a simple model taking into account the elasticity of the nanowire and its hydrodynamic interaction with the fluid medium captures the essential physics and  provides quantitative agreement with experimental measurements. 
We first solve for the detailed shape of the silver filament,  we  then predict the propulsion speed, and finally we compare our results with the experimental measurements.
Theoretical modelling of this type belongs to the general class of elastohydrodynamical problems, which has recently received a lot of attention in the literature \cite{machin58, wiggins98a, wiggins98b, lowe, powers02, yu, lauga07, coq}. 

Under our theoretical framework, we model the magnetic Ni segment as a rigid slender rod (radius $a_m = 100$nm, length  $L_m$=1.8$\mu$m) (the short Au segment is considered to be part of the rigid rod in this model, as discussed above, see Fig.~\ref{fig:setup}), and the flexible Ag nanowire (radius $a=50$nm, length $L=4\mu$m) as a classical Euler-Bernoulli beam \cite{landau}. We then employ a local fluid drag model, known as resistive force theory \cite{lauga2}, to describe the fluid-body interaction. The use of a local and linear theory significantly simplifies the analysis and is expected to provide quantitative agreement because geometric nonlinearities and nonlocal hydrodynamic effects were proven to be subdominant for gentle distortions of a slender body in previous work \cite{wiggins98a, powers02,yu}.

Notation for the model is shown in Fig.~\ref{fig:setup}(a). 
The external magnetic field precesses about the $z$-axis in the clock-wise direction, and can be described as $ \H = ( H_1 \cos \Omega t, -H_1 \sin \Omega t, H_0) = H_0 ( \h \cos \Omega t, -\h \sin \Omega t, 1) $, where $\h = H_1/H_0$ is the dimensionless relative strength of the rotating ($H_1$) and constant ($H_0$) components of the magnetic field. We study the regime where the nanowire follows synchronously the precessing magnetic field, rotating at the same angular frequency ($\Omega$) as the magnetic field about the $z$-axis. In addition, we can move in a rotating frame in which the magnetic field is fixed and the shape of the flexible nanowire does not change with time. In this frame, the precessing magnetic field is given by $\H = H_0 ( \h, 0, 1)$, and the nanowire has a non-changing shape $\r(s) = (\r_{\perp}(s), z(s)) = (x(s), y(s),z(s))$ in a background flow, $\vb$, rotating counter-clockwise about the $z$-axis: $ \vb = \Omega \mathbf{e}_z \times \mathbf{r}_{\perp}= \Omega (-y, x,0)$, where $\mathbf{e}_z$ is the unit vector in the $z$-direction and $s$ is the arclength parameter along the filament.

\subsection{Elastohydrodynamics at low Reynolds number}

We describe the fluid-body interaction by resistive force theory, which states that the local fluid drag depends only on the local velocity of the filament relative to the background fluid (although in a non-isotropic fashion). This is thus a local drag model which ignores hydrodynamic interactions between distinct parts of the filament, but was shown to be quantitatively correct for gentle distortions of the filament shape \cite{wiggins98a, powers02,yu}.  
The viscous force acting on the filament is thus expressed as
\begin{align}
\f_{\text{vis}} = -[ \xi_{\parallel} \t\t+\xi_{\perp} (1-\t\t)] \cdot \u,
\end{align}
where $\t(s)$ is the local tangent vector, $\u(s) = \U-\vb$ is the local velocity of the filament relative to the background flow $\vb$, and $\U$ is the swimming velocity of the nanomotor. Here, $\xi_{\parallel}$ and $\xi_{\perp}$ are the tangential and normal drag coefficients of a slender filament ($L \gg a$) and are given approximately by
\begin{align}\label{dragcoef}
\xi_{\parallel} = \frac{2 \pi \mu}{\log(L/a)-1/2}, \ \xi_{\perp} = \frac{4\pi\mu}{\log(L/a)+1/2},
\end{align}
where $\mu$ is the viscosity of the fluid (water, $\mu = 10^{-3}\text{Ns/m}^2$). Since the Ni and Ag segments have different aspect ratios ($L_m / a_m$ for the Ni segment), a different set of drag coefficients ( $\xi_{m\parallel}, \ \xi_{m\perp}$) is used for the rigid segment.

When the flexible Ag filament of the nanomotor is deformed, elastic bending forces arise trying to minimize the bending energy. This elastic bending force can be obtained by taking a variational derivative of the energy functional $\mathcal{E}=\frac{1}{2}\int_0^L A \left(\partial^2 \r / \partial s^2 \right)^2 ds$, where $A$ is the bending stiffness of the material. The elastic bending force is then given by
\begin{align}
 \f_{\text{elastic}} = - A \frac{\partial^4 \r}{\partial s^4} \cdot
\end{align}

Since we are in the low Reynolds number regime, inertial forces are negligible, and the local viscous fluid forces balance the elastic bending forces, $\f_{\text{vis}} + \f_{\text{elastic}}=\mathbf{0}$,
which yields the equation governing the filament elastohydrodynamics
\begin{align}
[ \xi_{\parallel} \t\t+\xi_{\perp} (1-\t\t)] \cdot \u = - A \frac{\partial^4 \r}{\partial s^4} \cdot
\end{align}

The flexible Ag filament is clamped to the magnetic Ni segment, which is assumed to be rigid and straight. Hence, its position vector is given by $\r_m (s) = \r\mid_{s=L} + \t\mid_{s=L} (s-L)$, where $s \in [L,L+L_m]$. 

\subsection{Nondimensionalization}
We now nondimensionalize the variables and equations and identify the relevant dimensionless parameters governing the physics of this problem. Specifically, we scale lengths by $L$, rotation rates by $\Omega = 2\pi f$, times by $\Omega^{-1}$, velocities by $L \Omega$, fluid forces by $\xi_{\perp} L^2 \Omega$, fluid torques by $\xi_{\perp} L^3 \Omega$, elastic forces by $A/L^2$, and elastic torques by $A/L$. Using the same symbols for simplicity, the dimensionless elastohydrodynamic equation now reads
\begin{align}\label{nodim}
[ \gamma^{-1} \t\t+(1-\t\t)] \cdot \u = - \Sp^{-4} \frac{\partial^4 \r}{\partial s^4},
\end{align}
where we have defined $\gamma = \xi_{\perp}/\xi_{\parallel}$, and $\Sp = L \left( \xi_{\perp} \Omega / A\right)^{1/4} $ is termed the sperm number, which characterizes the relative influence of the fluid and bending forces.

\subsection{Asymptotic analysis}
The geometrical nonlinearity of Eq.~\eqref{nodim} renders the elastohydrodynamic equation only solvable via numerical simulation in most situations.  Here we are able to illustrate the essential physics of flexible nanomotor propulsion analytically via an asymptotic analysis for the case where $\h=H_1/H_0$ is small. Such an approximation drops the geometrical nonlinearities and, as will be shown below, separates the task of determining the filament shape and swimming velocities of the nanomotor, as the axial velocities are one order of magnitude smaller than the transverse velocities,  the axial swimming kinematics being thus slaved to the transverse kinematics \cite{lauga07}. Even with this simple model, we find that the theoretical predictions agree well with the experimental measurements. In the experiments, we do not observe very significant distortion of the flexible Ag filament, which might explain the success of this simple model.

As the nanomotor was observed to propel unidirectionally in the $z$-direction in the experiments (i.e.~the direction about which the actuating magnetic field precesses), we write the swimming speed as $\U = (0,0,U)$ and aim at predicting the leading order swimming speed in $\h$. We do not expect any $O(\h^0)$ deformation nor swimming velocities, and hence  the appropriate expansions for the deformation of the nanowire and the swimming speed are given by
\begin{align}
\r_{\perp}(z) &= \h \ \r_{\perp_{1}}(z) + \h^2 \r_{\perp_{2}}(z) + O(\h^3),\\
U &= \h \ U_1 + \h^2  \ U_2 + O(\h^3),
\end{align}
where we have $s \approx z + O(\h^2)$.

\begin{figure*}
\centering
\includegraphics[width=0.9\textwidth]{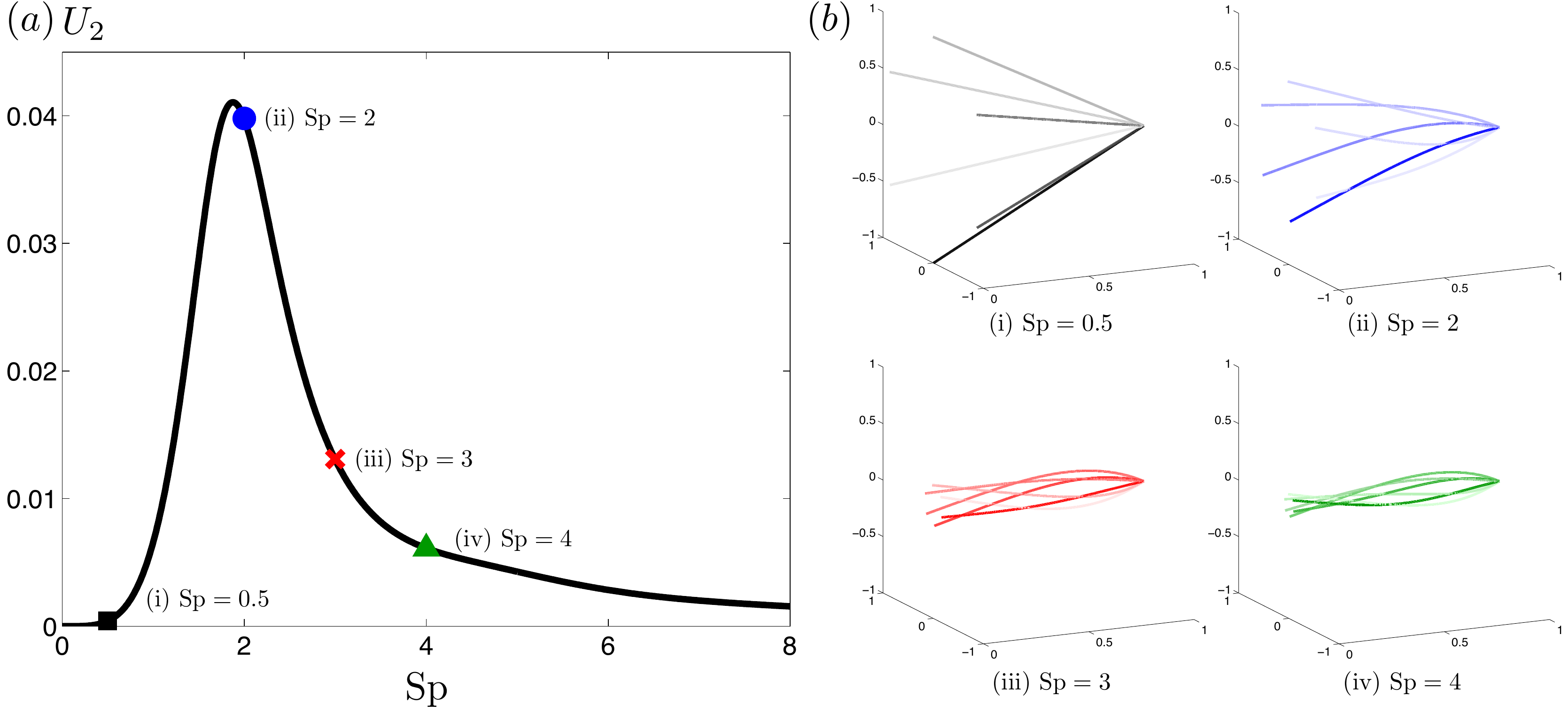}
\caption{\label{fig:SpermNumber}(a) Variation of the dimensionless propulsion speed  at second order, $U_2$,  with  the sperm number, $\Sp$. (b) Superimposed snapshots of predicted three-dimensional shape of the Ag nanowire at equal time intervals ($t=[T/6, 2T/6, ..., 5T/6, T]$ from dark to bright color, where $T$ is the period of the rotating magnetic field),  for four different sperm numbers. The Ni head is not shown here for simplicity.}
\end{figure*}

The elastoydrodynamic equation is a fourth-order partial differential equation in space, and needs thus to be supplied with four boundary conditions. We prescribe dynamic boundary conditions at the free end $z=0$, requiring it to be force-free and torque-free, which is $\partial^3 \r/\partial z^3 (z=0)=\mathbf{0}$ and $\partial^2 \r/\partial z^2 (z=0)=\mathbf{0}$ respectively. Since the deformed shape rotates about the $z$-direction, without loss of generality, we assume the Ni head lies on the $x-z$ plane. We then prescribe kinematic boundary conditions at the other end $z=1$: $\mathbf{r}_{\perp} (z=1) = (b,0)$ and $\partial \mathbf{r}_{\perp}/\partial z (z=1)= (\text{h},0)$. From experimental observations, the value of $b$ is seen to be negligibly small ($b \approx 0$) and is difficult to measure accurately. Here, for simplicity, we thus take $b= 0$ in our calculations below. In this geometric model, we assume that the slope of the magnetic Ni head, $\partial x(z)/ \partial z (z=1)$, follows the slope of the external field ($\h=H_1/H_0$), which is a good approximation when the magnetic field strength is strong or when the frequency of actuation is low, such that the Ni head can align closely with the external magnetic field. The magnitude of the magnetic torque can be compared with the viscous torque acting on the Ni head, and their ratio is given by the so-called Mason number, \text{Ma}. The ratio varies from 0.018 -- 0.12, for frequency varying form 5Hz to 35Hz. One can also compare the magnetic torque to the characteristic viscous torque acting on the Ag filament, and it varies from 0.13 -- 0.93, for the same range of frequency. In both cases, \text{Ma} is thus typically small and is at most $O(1)$ at high frequencies. Therefore, within the range of frequency explored in the experiment, our geometrical model is considered to be a valid approximation. At higher frequencies,  we would get $\text{Ma}\gg1$, which  would play a role in the boundary condition at $z=1$. In that regime, the viscous torque would dominate the typical actuation torque by the magnetic field, and the Ni segment would therefore not be able to align with the magnetic field closely. We expect that the slope of the Ni rod might then be smaller than that of the magnetic field, and the phase lag between the motion of the Ni segment and the magnetic field could be substantial. As a result, a degradation in the propulsion performance would be expected to occur in this regime.

\subsubsection{Determining the flexible filament shape: O($\h$) calculations}
~At order $O(\h)$, the local viscous force is given by $\f_{\text{vis}} = \h \ (-y_1(z), x_1(z), -\gamma^{-1} U_1) + O(\h^2)$. From here, 
we can integrate the $O(\h)$ local viscous force in the $z$-direction over the entire nanomotor and since this total force needs to vanish because of the absence of external forces, we find $U_1=0$:  swimming occurs therefore at order $O(\h^2)$. 
The elastic force is given by $\f_{\text{elastic}}= - \h \ \Sp^{-4} (\partial^4 x_1/\partial z^4, \partial^4 y_1/\partial z^4,0)+O(\h^2)$. Balancing the local viscous and elastic forces in the transverse directions yield the hyper-diffusion equations \cite{powers02}
\begin{align}
-y_1 &= \Sp^{-4} \frac{\partial^4 x_1}{\partial z^4},\label{eqn:hyperx}\\
x_1 &=\Sp^{-4} \frac{\partial^4 y_1}{\partial z^4}\label{eqn:hypery},
\end{align}
which govern the first order filament shape. The general solution to this system of partial differential equations is given by 
\begin{align}
x_1(z) &= \sum_{n=1}^{8} A_n \exp \left( \Sp \ r_n z \right),\label{eqn:solutionx}\\
y_1(z) &= \sum_{n=1}^{8} -A_n r^4_n\exp \left(\Sp \ r_n z \right),\label{eqn:solutiony}
\end{align}
where $r_n$ is the $n$-th eight roots of $-1$, and $A_n$ are complex constants to be determined by the boundary conditions. The boundary conditions to this order at $z=0$ are given by $\partial^3 x_1/\partial z^3 (z=0) = \partial^3 y_1/\partial z^3 (z=0)=\partial^2 x_1/\partial z^2(z=0) =\partial^2 y_1/\partial z^2 (z=0) =0$. The appropriate boundary conditions at $z=1$ are given by $ x_1(z=1) = y_1(z=1)=0$, $\partial x_1/\partial z (z=1)=1$, and $\partial y_1/\partial z (z=1)=0$. The $O(\h)$ filament shape is now completely determined.

\subsubsection{Determining the swimming speed: O($\h^2$) calculations}
~At order $O(\h^2)$, the local viscous fluid force acting on the flexible filament in the $z$-direction is given by  
\small
$$
\mathbf{e}_z \cdot \f_{\text{vis}_2}= \left\{ \begin{array}{rl}
 -(\gamma^{-1}-1) \Lambda(z)-\gamma^{-1} U_2 &\mbox{, \,\,$0 \le z < 1$}, \\
\frac{\xi_{m\perp}}{\xi_{\perp}} \left[ -(\gamma_m^{-1}-1) \Lambda(z)-\gamma_m^{-1} U_2 \right]  & \mbox{,\,\, $1<z \le 1+l_m$},
       \end{array} \right.
$$
\normalsize 
where $l_m = L_m/L$, and we have introduced the function $\Lambda(z) =  y_1(z) \partial x_1/ \partial z (z)-x_1(z) \partial y_1/\partial z (z) $.  Since the nanomotor is overall force-free, the second order swimming speed $U_2$ can be determined by integrating the local viscous fluid in the $z$-direction over the entire nanomotor and requiring this total force to vanish, i.e.
\begin{align}
\int_0^{1+l_m} \mathbf{e}_z \cdot \f_{\text{vis}_2} dz = 0,
\end{align}
and we see that the swimming speed is slaved to the first order filament shape $(x_1(z), y_1(z))$ via the function $\Lambda(z)$. Upon simplification with Eqs.~\eqref{eqn:hyperx} and \eqref{eqn:hypery} and the boundary conditions at $z=0$, we obtain 
\small
\begin{align}\label{eqn:speed}
U_2 = & \frac{ 1-\gamma}{\Sp^{4}(1+\alpha l_m)} \times \\ \notag
&\biggl[ \frac{\partial x_1}{\partial z}\frac{\partial^3 x_1}{\partial z^3}-\frac{1}{2} \left(\frac{\partial^2 x_1}{\partial z^2}\right)^2+\frac{\partial y_1}{\partial z}\frac{\partial^3 y_1}{\partial z^3}-\frac{1}{2} \left(\frac{\partial^2 y_1}{\partial z^2}\right)^2 \biggr] _{z=1},
\end{align}
\normalsize 
where $\alpha = \xi_{m\parallel}/\xi_{\parallel}$. In dimensional form, the leading order swimming speed, is given by
\small
\begin{align}\label{eqn:speed}
U = \ & \h^2 \frac{A(\xi_{\parallel}-\xi_{\perp})}{\xi_{\perp}(L \xi_{\parallel}+L_m \xi_{m\parallel})} \times \\ \notag
&\bigg[ \frac{\partial x_1}{\partial z}\frac{\partial^3 x_1}{\partial z^3}-\frac{1}{2} \left(\frac{\partial^2 x_1}{\partial z^2}\right)^2+\frac{\partial y_1}{\partial z}\frac{\partial^3 y_1}{\partial z^3}-\frac{1}{2} \left(\frac{\partial^2 y_1}{\partial z^2}\right)^2 \biggr]_{z=1}\\
&+O(\h^3)\nonumber.
\end{align}
\normalsize 
As in previous work \cite{lauga07,yu,coq,coq09,gray55,becker}, we observe  that this mode of propulsion relies on the drag anisotropy of slender filaments, $\gamma = \xi_{\perp}/\xi_{\parallel}\neq1$. Indeed,  when $\gamma=1$, the swimming speed vanishes. Note that for very slender filaments, we have $\gamma \approx 2$ 
(see Eq.~\ref{dragcoef}).  

We also see that the swimming speed scales quadratically with the relative strength of the rotating and constant components of the magnetic field, $U \sim \h^2$, for $\h \ll 1$. This scaling is confirmed by a complementary asymptotic calculation valid for  low sperm numbers (see Appendix for details).

\begin{figure*}[!]
\centering
\includegraphics[width=0.9\textwidth]{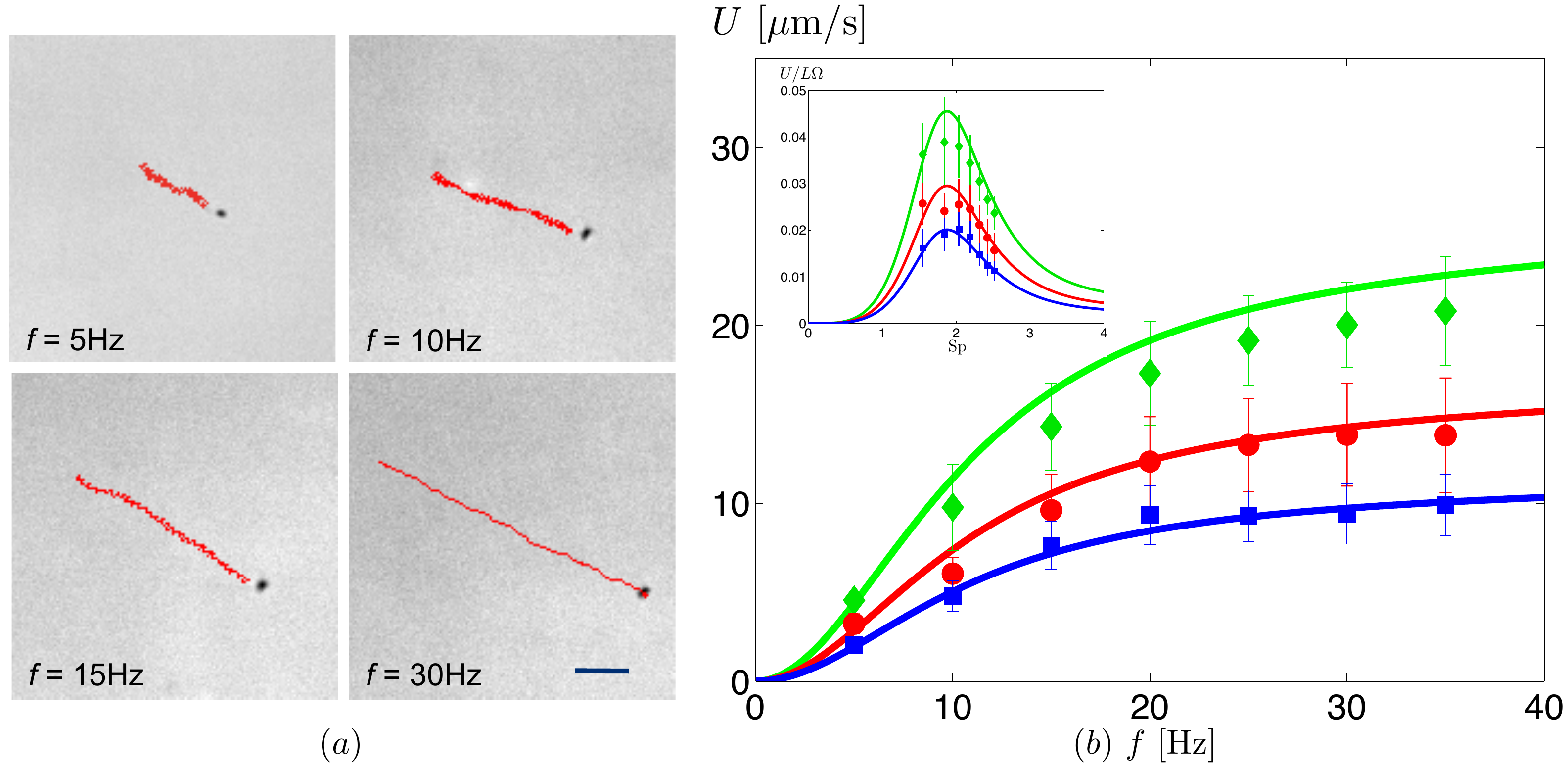}
\caption{\label{fig:TE}Dependence of the nanomotor swimming speed on the actuation frequency. (a) Superimposed  trajectories of the same Ni-Ag nanomotor at different frequencies $f= 5$, 10, 15, 30Hz (as indicated) over a 3-second period (red lines), with $H_1 = 10$G and $H_0 =9.5$G. The scale bar is 10$\mu$m.  (b) Speed-frequency characteristics of flexible nanowire motors. Symbols represent experimental data for different setups of the magnetic fields: blue squares ($H_1$ = 10G, $H_0$ = 9.5G); red circles ($H_1$ = 10G, $H_0$ = 11.8G); green diamonds ($H_1$ = 10G, $H_0$ = 14.3G). Error bars show standard deviations of the measured speeds (20 samples). The solids lines show the theoretical predictions (Eq.~\ref{eqn:speed}) with $A = 3.6 \times 10^{-24}$Nm$^2$. The inset in (b) displays the dependence of the swimming speed  on the sperm number, $\Sp$.}
\end{figure*}

Next, we plot our predicted dimensionless second order swimming speed as a function of the sperm number $\Sp$ (Fig.~\ref{fig:SpermNumber}a) together with the predicted filament shapes (Fig.~\ref{fig:SpermNumber}b), and observe three different characteristic regimes. The sperm number $\Sp$ is the most important dimensionless group governing the propulsion performance. For $\Sp \ll 1$, bending forces dominate and the filament is effectively straight (Fig.~\ref{fig:SpermNumber}b-i).  Hence,   the filament motion is almost kinematically reversible, and it produces small  propulsion. Quantitatively, a small $\Sp$ asymptotic analysis presented in the Appendix reveals that the dimensionless swimming speed grows with the fourth power of the sperm number, $U_2 \sim \Sp^4$, for $\Sp \ll 1$. On the other hand, from Eqs.~\eqref{eqn:solutionx} and \eqref{eqn:solutiony}, we see that most deflection is concentrated around a small region $0\le z<1/\Sp$, when $\Sp \gg 1$, due to the exponential decay of the solution amplitude. In this regime ($\Sp \gg 1$), the viscous forces dominate, and propulsion is inefficient because a large portion of the filament has small deflection and thus experiences drag but contributes  to no  thrust (Figs.~\ref{fig:SpermNumber}b-iii \& iv). As a result, we expect optimal swimming to occur when $\Sp $ is of order one, where the total drag of the nanomotor is kept low while the drag-induced bending is fully exploited to produce propulsion. This is confirmed in our calculation, and we observe the optimal sperm number to occur at $\Sp\approx 2$, which gives a maximum propulsion speed of $U_2 \approx 0.042$  (Fig.~\ref{fig:SpermNumber}a). The filament shape close to optimal swimming ($\Sp=2$) is shown in Fig.~\ref{fig:SpermNumber}b-ii.

\subsection{Comparison with experimental measurements}

Under fixed magnitude of the  rotating and constant components of the magnetic field, the swimming speed of a nanomotor was measured with the frequency of the magnetic field varying between 0 to 35Hz. The experiment was repeated on the same nanomotor for three different settings of magnetic field strengths (shown using three different symbols with error bars in Fig.~\ref{fig:TE}b; a total of 20 different nanowires were sampled). The rotating magnetic field strength $H_1$ was kept constant at $H_1 = 10$G, and the constant magnetic field strength was set to be $H_0=14.3$G (blue squares), $H_0=11.8$G (red circles), and $H_0=9.5$G (green diamonds). 

We then compare in Fig.~\ref{fig:TE}b our theoretical predictions (solid lines) with experimental measurements, plotted as swimming velocity vs.~frequency (main figure) or sperm number (inset).  In our theoretical model, the value of the bending stiffness $A$ of the flexible filament is unknown. Standard bending stiffness of pure silver is inapplicable here since the dissolution of silver in hydrogen peroxide rendered the filament a porous structure with significantly reduced strength and a different chemical composition ($\text{Ag}_2 \text{O}$, AgOH).  A value of $A = 3.6 \times 10^{-24}\text{Nm}^{-2}$ fits, with the least total squared errors, the experimental data with $H_1=10\text{G}, H_0 = 14.3\text{G}$ (blue squares), which is the case where our model is expected to work best as the ratio $\h = H_1/ H_0$ is the smallest. This bending stiffness is then used to predict the propulsion speed under different magnetic field settings (green and red solid lines in Fig.~\ref{fig:TE}b, see captions for details).

The theoretical model is seen to capture both qualitatively and quantitatively the speed-frequency characteristics of these flexible nanomotors. Qualitatively, the rate of change of the swimming speed with respect to the frequency increases at low frequencies ($U \sim f^2$ for small $f$, as shown in the Appendix), but then gradually decreases as the frequency continues to increase, and eventually levels off at high frequencies. Physically, increasing the actuation frequency is equivalent to increasing the sperm number. When the frequency is varied from 0 to 35Hz, it corresponds to a variation of the sperm number $\Sp$ from 0 to 2.6, experiencing a degradation in swimming performance beyond the optimal sperm number $\Sp \approx 2$, which corresponds to a frequency of around 15Hz in our experiment. This degradation manifests as a less-than linear speed-frequency variation (since the dimensional swimming speed scales as $L\Omega$, linearly in $\Omega$) beyond the frequency $15$Hz, resulting in the level-off at higher frequencies. As noted above, at very high frequencies, the magnetic Ni head will be unable to follow synchronously the rapid rotating magnetic field. The dynamics of propulsion will be more complicated in that regime, and the simple model presented here will likely be inapplicable.

The agreement between our theoretical model  and our experimental results is very satisfactory. The discrepancies are larger for the setup $H_1=10\text{G}, H_0 = 9.5\text{G}$ (green lines and squares), which is expected because $\h \approx 1.1$ in this case and the asymptotic assumption of small $\h$ is less valid. Note that our measurements did not sample the low $\Sp$ regime as  in our experiments, swimming at low frequencies appear to be significantly influenced  by Brownian motion.

Our model has only one fitting parameter, the bending stiffness $A$, which -- as explained above -- we fit to the bottom data set in Fig.~\ref{fig:TE}b, and use to predict the other two data sets. The estimated value we obtain from the fitting is equivalent to a pure silver  nanowire of diameter $\approx 6$nm (with elastic modulus, E $= 80$GPa), which is much smaller than the diameter of flexible segment observed. This is expected because the chemical composition of silver is altered after the dissolution, and a large portion of the flexible nanowire is indeed a thick layer of surface byproducts formed after the chemical reaction, which contributes little, if any, to the bending strength. The diameter of the actual structural filament that bears the bending loads is difficult to measure experimentally (see details of the structure in 
Fig.~\ref{fig:setup}b). In addition, non-uniform chemical reactions lead to strong local defects or points of weakness along the nanowire, which might significantly reduce the bending strength. We can compare our estimated bending stiffness, $A = 3.6 \times 10^{-24}\text{Nm}^2$, with the bending stiffness of typical flagella of natural microscopic swimmers, such as eukaryotic spermatozoa, which also rely on the flexibility of flagella for propulsion. These biological filaments have their bending stiffnesses ranging from  $10^{-24}\text{Nm}^2$ \cite{fujime} to $10^{-22}\text{Nm}^2$ \cite{hoshikawa}, which is the range in which our estimated value lies.

\section{Discussion}
\begin{figure}[t]
\centering
\includegraphics[width=0.48\textwidth]{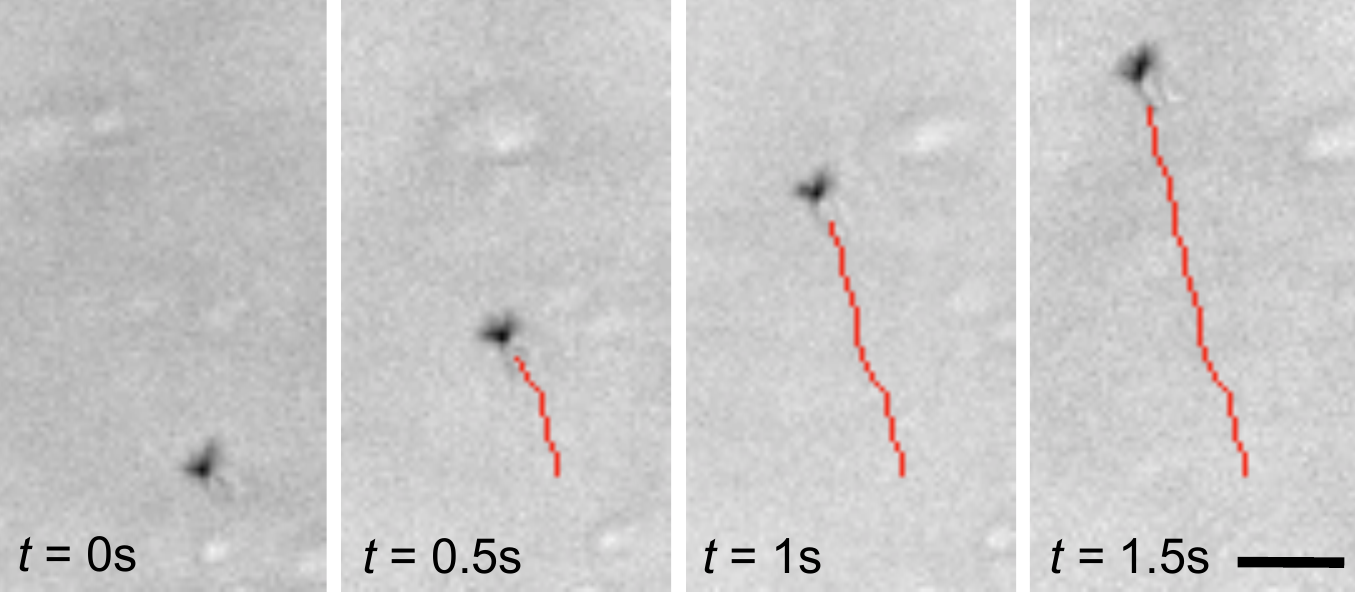}
\caption{\label{fig:serum}Time lapse images (time as indicated) of the motion of nanowire motor (velocity, $U = 15\mu$m/s) in human serum at $f = 15$Hz, with $H_1 =10$G and $H_0 = 9.5$G. Scale bar is 5$\mu$m.}
\end{figure}

In this work, we designed and fabricated a high-speed fuel-free nanomotor utilizing the flexibility of nanowires for propulsion. These flexible nanomotors demonstrate a number of 
 advantages: first, the fabrication process is relatively simple and involves a common template-directed electrodeposition protocol of nanowires; second, these nanowire motors are able to propel at  high speeds, both dimensional (up to $\approx 21 \mu$m/s) and dimensionless (up to 0.164 body lengths per revolution), and their performance compares very well with natural microorganisms and other synthetic locomotive systems; third, they are actuated by external magnetic field and do not require specific chemical environments  (fuels) for propulsion, which is preferable for biomedical applications. Indeed, the performance of the nanomotors reported here is not affected by the presence of ions or other  chemical species, and they are able to propel equally well in real biological settings. As an illustration, we have placed these flexible nanomotors in human serum, and observed similar propulsion behaviours (Fig.~\ref{fig:serum} and Video 3\dag). This demonstrates an exciting potential of these flexible nanomotors for future biomedical applications such as targeted drug delivery systems, or cell manipulation. 

The fundamental physics of the flexible nanomotors has been illustrated by a simple analytical elastohydrodynamic model. The propulsion characteristics were experimentally studied and compared with the theoretical model, with good agreement. Strictly speaking, the results of the asymptotic model presented in this paper are valid only for $\h \ll 1$. However, as shown in other previous studies which compared asymptotic results with numerical and experimental studies \cite{wiggins98a, powers02,yu}, these asymptotic models often remain valid even up to $\h \sim 1$, meaning that geometrical nonlinearities do not play very significant roles. Our results also ignore the hydrodynamic effect of the bottom surface close to which the nanomotors are propelling. As we estimated experimentally the distance of the filaments to the surface to be on the order of microns, and therefore on the order of the swimmer size, we do not expect strong hydrodynamic effects from the surface, which might explain the success of our simple modelling approach. Further progress in theoretical modelling most likely have to be obtained numerically. More accurate yet complicated descriptions of the hydrodynamic interactions can be achieved using methods such as slender body theory \cite{tornberg}, or regularized Stokeslets \cite{cortez}. Tension, self-spinning, and twist strains of the filament may also be considered for improvements.

\section{Acknowledgements}
We thank Professor Eric Fullerton, Erik Shipton, and Daniel Kagan for their help on the magnetic setup, and Allen Pei, Adam Ponedal for assisting in the nanowire preparation. Useful discussions with Dr.~Saverio Spagnolie are acknowledged. Funding by the National Science Foundation (Grant Nos. CBET-0746285 to E.~L. and CBET-0853375 to J.~W.), and the Croucher Foundation (through a scholarship to O.~S.~P.) is gratefully acknowledged.

\appendix

\section{Swimming at low sperm numbers}
In this appendix, we consider another physically interesting asymptotic limit, the low sperm number limit, $\Gamma = \Sp^4 \ll1$. The results in this asymptotic limit are not expected to provide quantitative agreement with the experimental measurements, since the value of $\Gamma$ in the experiment is typically large. Nevertheless, this analysis still allows us to reveal different scaling behaviours of the propulsion speed at low sperm numbers. Keaveny and Maxey \cite{keaveny} investigated the propulsion of a flexible filament with distributed magnetic actuation. With a resistive force model, they considered the low sperm number limit and found that the propulsion speed scales with the fourth power of the sperm number, $U \sim \Sp^4$, at low sperm numbers. We will follow closely their method of solution and perform similar calculations here to show that the same scaling holds for our flexible nanowire motor subject to boundary actuation. An explicit formula for the leading order propulsion speed in $\Gamma$ will be derived. Expanding this formula for small $\h$ will confirm our scaling of $U$ with $\h$ from the small-$\h$ asymptotic analysis in the main text.

First, the problem is formulated below, taking into account the effects of twisting, self-spinning, and inextensibility of the nanowire. Denoting  $\N(s,t)$ and $\M(s,t)$ as the resultant internal force and moment on a cross section, the local force and moment balances are
\begin{align}
\frac{\partial \N}{\partial s} &= (\xi_{\parallel}-\xi_{\perp}) (\t \cdot \u) \t+ \xi_{\perp}\u, \label{eqn:full1}\\
\frac{\partial \M}{\partial s} + \t \times \N &= \xi_{r} a^2 (\OOmega \cdot \t) \t,
\end{align}
where $\OOmega(s,t)$ is the angular velocity, and $\xi_{r}=4\pi \mu$ is the resistive coefficient for the viscous torque produced by self-spinning (rotation about its own local axis, $\t$) of the filament. The internal moment $\M(s,t)$ has a constitutive relation
\begin{align}
\M = A \t \times \frac{\partial \t}{\partial s} + K_t \frac{\partial \Psi}{\partial s} \t,
\end{align}
where $K_t$ is the twist modulus of the filament and $\Psi(s,t)$ is the twist angle. In contrast to the propeller studied by
Keaveny and Maxey \cite{keaveny}, the magnetic torque does not come into the local moment balance in our case, but only through the boundary condition. The boundary conditions are given by the balance of external forces and torques at the ends of the flexible filament. We have a free end at $s=0$. The external forces and torques at $s=1$ are given by the total viscous force and viscous torque together with the magnetic torque acting on the Ni segment, which is modelled as a slender rigid rod: 
\begin{align}
\N(s=0) &= \mathbf{0},\\
\M(s=0) &= \mathbf{0},\\
\N(s=1) &= - L_m \left[ (\xi_{m\parallel}-\xi_{m\perp}) (\t \cdot \u) \t +\xi_{m\perp}\u \right]_{s=1} \notag \\ 
& \ \ \ \ + \xi_{m\perp} \frac{L_m^2}{2}\OOmega \times \t \mid_{s=1} ,\\
\M(s=1) &= M \t \mid_{s=1} \times \H \notag \\ 
& \ \ \  \ -\xi_{m\perp}\left\{ \frac{L_m^2}{2} \t \times \u+ \frac{L_m^3}{3}[\OOmega-(\t \cdot \OOmega) \t]  \right\}_{s=1},
\end{align}
where $M=M_s a_m^2 \pi L_m$ is the strength of the magnetic moment of the Ni segment and $M_s = 485 \times 10^{3}$A/m is the spontaneous magnetization of Ni. 

To study the low sperm number limit, we adopt the following nondimensionalizations: we scale times with $\Omega^{-1}$, lengths with $L$, $\H$ with $H_0$, elastic forces with $A/L^2$, and elastic torques with $A/L$. With these nondimensionalizations, the dimensionless equations (using the same variables for simplicity) now read
\begin{align}
\frac{\partial \N}{\partial s} &= \Gamma \left[ (\gamma^{-1}-1) (\t \cdot \u) \t+ \u \right], \label{eqn:full1}\\
R \Gamma (\OOmega \cdot \t) \t &= \t \times \frac{\partial^2 \t}{\partial s^2} + \t \times \N + K \frac{\partial}{\partial s} \left( \frac{\partial \Psi}{\partial s} \t \right),
\end{align}
where $R = \xi_{r} a^2/\xi_{\perp} L^2$, $K=K_t/A$, and $\Gamma = \Sp^4$. The dimensionless boundary conditions are
\begin{align}
\N(s=0) &= \mathbf{0},\\
\M(s=0) &= \mathbf{0},\\
\N(s=1) &= -\Gamma \beta \times \left\{ \left[ (\gamma_m^{-1}-1) (\t \cdot \u) \t+\u \right] l_m+ \frac{l_m^2}{2}\OOmega \times \t \right\}_{s=1},\\
\M(s=1) &= C_m \t \mid_{s=1}\times \H -\Gamma \beta \left\{ \frac{l_m^2}{2} \t \times \u+ \frac{l_m^3}{3}\left[\OOmega-(\t \cdot \OOmega) \t \right] \right\}_{s=1},
\end{align}
where $l_m = L_m/L$, $\beta =\xi_{m\perp}/\xi_{\perp}$, $\gamma_m = \xi_{m\perp}/\xi_{m\parallel}$, and $C_m = M_s a_m^2 \pi L_m H_0 L /A$ is a dimensionless parameter characterizing the relative strength of the magnetic and elastic torques. Finally, we have the inextensibility condition
\begin{align}
\t(s,t) \cdot \t(s,t) =1.\label{eqn:full2}
\end{align}

Eqs.~\eqref{eqn:full1} through \eqref{eqn:full2} completely describe the full swimming problem (within the realm of resistive force theory and classical elastic beam theory) without making any assumption. There is no restriction on the validity of the solution to this system, but the solution has to be obtained numerically, with special attention on the nonlinearities arising in the differential equation and the boundary conditions. To make analytical progresses, we consider the asymptotic limit $\Gamma \ll 1$, and calculate the leading order swimming speed in $\Gamma$. Following the method and notations by Keaveny and Maxey \cite{keaveny}, we assume the filament attains a constant shape at steady-state and rotates about the $z$-axis synchronously with the external magnetic field, hence we write the dimensionless steady-state conformation of the filament as
\begin{align}
x(s,t) &= -b(s) \cos[t+\phi(s)],\\
y(s,t) &= -b(s) \sin[t+\phi(s)],\\
z(s,t) &= \alpha (s) + \tilde{U} t,
\end{align}
where $\tilde{U} = U/L\Omega$ is the dimensionless swimming speed in the $z$-direction, whereas $\alpha(s)$,  $b(s)$, and $\phi(s)$ are geometrical unknowns to be determined. Like the small-$\h$ asymptotic analysis in the main text and in Ref.~\cite{keaveny}, here we have considered unidirectional swimming in the $z$-direction. In addition, since the functions $\alpha(s), b(s)$ and $\phi(s)$ are independent of time, we only find the solution at one specific time, $t=0$ \cite{keaveny}. The magnetic field at $t=0$ is given by $\H = (\h, 0, 1)$. We seek expansions in $\Gamma$ in the form of
\begin{align}
\N &= \N_0 + \N_1 \Gamma + O(\Gamma^2),\\
\M &= \M_0 + \M_1 \Gamma + O(\Gamma^2),\\
\frac{d \alpha}{d s} &= \frac{d \alpha_0}{d s} +\frac{d \alpha_1}{d s} \Gamma + O(\Gamma^2),\\
\frac{d b}{d s} &= \frac{d b_0}{d s} +\frac{d b_1}{d s} \Gamma + O(\Gamma^2),\\
\frac{d \phi}{d s} &= \frac{d \phi_1}{d s} \Gamma + O(\Gamma^2),\\
\frac{d \psi}{d s} &= \frac{d \psi_1}{d s} \Gamma + O(\Gamma^2),\\
\tilde{U} &= \tilde{U}_0+\tilde{U}_1 \Gamma + O(\Gamma^2), 
\end{align}
and similar expansions hold for other variables. With these expansions, we can express the local tangent $\t(s)$ and velocity $\u(s,t)$ as
\begin{align}
\t(s) &= \t_0 + \t_1 \Gamma + O(\Gamma^2),\\
&= \left (-\frac{d b_0}{d s},0, \frac{d \alpha_0}{d s} \right) \\ \notag
& \ \ \ \ + \left (-\frac{d b_1}{ds}, -\frac{d \left(b_0 \phi_1\right)}{ds}, \frac{d \alpha_1}{d s}\right) \Gamma + O(\Gamma^2),\\
\u(s) &= \left( 0, -b_0 , \tilde{U}_0 \right) + \left( b_0 \phi_1, -b_1, \tilde{U}_1\right) \Gamma + O(\Gamma^2).
\end{align}
In the following section, we will perform the calculations order by order.

\subsection{$O(\Gamma^0)$ calculations}
The $O(\Gamma^0)$ local balance of forces and torques are given by
\begin{align}
\frac{d \N_0}{d s} &=\mathbf{0},\\
\mathbf{0} &= \t_0 \times \N_0 + \t_0 \times \frac{d^2 \t_0}{d s^2} + K \frac{d}{d s} \left( \frac{d \Psi_0}{d s}\t_0\right),
\end{align}
with boundary conditions
\begin{align}
\N_0 (s=0) &=\mathbf{0}, \\
\M_0 (s=0) &=\mathbf{0}, \\
 \N_0(s=1) &=\mathbf{0},\\
 \M_0(s=1) &= C_m \t_0 \mid_{s=1} \times \H,
\end{align}
and inextensibility condition
\begin{align}
\left( \frac{d \alpha_0}{ds} \right)^2 +\left( \frac{d b_0}{ds} \right)^2 =1.
\end{align}

The solution at this order is given by
\begin{align}
\N_0(s) &= \mathbf{0},\\
\frac{d \alpha_0}{ d s} &= \frac{1}{\sqrt{1+\h^2}},\\
\frac{d b_0}{ d s} &= \frac{-\h}{\sqrt{1+\h^2}}, \label{eqn:b0}\\
\Psi_0(s) &= \tilde{U}_0 = 0.
\end{align}

\subsection{$O(\Gamma)$ calculations}
The $O(\Gamma)$ local balance of forces and torques are given by
\begin{align}
\frac{d \N_1}{d s} &= (\gamma^{-1}-1) (\t_0 \cdot \u_0)\t_0 +\u_0,\\
\mathbf{0} &= \t_0 \times \N_1 + \t_0 \times \frac{d^2 \t_1}{d s^2} + \t_1 \times \frac{d^2 \t_0}{d s^2} + K \frac{d^2 \Psi_1}{d s^2} \t_0,
\end{align}
with boundary conditions
\begin{align}
\N_1 (s=0) &= \mathbf{0},\\
\M_1 (s=0) &= \mathbf{0},\\
\N_1 (s=1) &= -\beta l_m \times  \notag \\
& \ \ \ \ \left[ (\gamma_m^{-1}-1) (\t_0 \cdot \u_0) \t_0 + \u_0 + \frac{l_m}{2} \OOmega_0 \times \t_0   \right]_{s=1},\\
\M_1 (s=1) &= C_m \t_1 \mid_{s=1} \times \H \notag \\
& \ \ \ \ - \beta l_m^2 \left\{  \frac{\t_0 \times \u_0}{2}+\frac{l_m}{3} \left[ \OOmega_0 -(\t_0 \cdot \OOmega_0) \t_0 \right]  \right\}_{s=1},
\end{align}
and the inextensibility condition
\begin{align}
\frac{d \alpha_1}{d s} = \h \frac{d b_1}{d s}\cdot
\end{align}

From the solution at $O(\Gamma^0)$, Eq.~\eqref{eqn:b0}, we can integrate to get $b_0 (s) = -\h s/\sqrt{1+\h^2}+C_1$, where $C_1$ is a constant to be determined. By satisfying the equations and boundary conditions at this order, we find that 
\begin{align}
C_1 &= \frac{\h\left[ 1+ \beta l_m (2+l_m)\right]}{2 \sqrt{1+\h^2} (1+\beta l_m )} \cdot
\end{align}
The solution at this order is then given by
\begin{align}
\frac{d (b_0 \phi_1)}{d s} &= \frac{\h s^4}{24 \sqrt{1+\h^2}} - \frac{C_1 s^3}{6}+B_1,\\
\frac{d b_1}{d s} &= \frac{d \alpha_1}{d s} = \frac{d \Psi_1}{d s} =0,
\end{align}
where
\begin{align}
B_1 = \ &\frac{\h}{24 C_m \left(1+\h^2\right) (1+ \beta l_m)} \times \notag \\
& \biggl\{ 2+C_m \sqrt{1+\h^2} \left[ 1+\beta l_m (3+2 l_m)  \right] \notag \\
& +2  \beta l_m \left[ 4+l_m (6+l_m (4+ \beta l_m )) \right] \biggr\}.
\end{align}

\subsection{$O(\Gamma^2)$ calculations}
The $O(\Gamma^2)$ local balance of forces and torques are given by
\begin{align}
\frac{d \N_2}{d s} &= (\gamma^{-1}-1) (\t_1 \cdot \u_0+\t_0 \cdot \u_1) \t_0 + \u_1,\label{eqn:G2F}\\
R (\OOmega_1 \cdot \t_0 +\OOmega_0 \cdot \t_1 ) \t_0 &= \t_0 \times \N_2 + \t_1 \times \N_1 \\ \notag
& \ \ \ \ + \t_0 \times \frac{d^2 \t_2}{d s^2} + \t_1 \times \frac{d^2 \t_1}{d s^2}+ K \frac{d^2 \Psi_2}{d s^2} \t_0,
\end{align}
with boundary conditions
\small
\begin{align}
\N_2 (s=0) &= \mathbf{0},\label{eqn:G2BC0}\\
\M_2 (s=0) &= \mathbf{0},\\
\N_2(s=1) &= - \beta l_m \biggl[ (\gamma_m^{-1}-1) (\t_0 \cdot \u_1 + \t_1 \cdot \u_0) \t_0 \notag \\
& \ \ \ \ +\u_1+\frac{l_m}{2} \left( \OOmega_0 \times \t_1 +\OOmega_1 \times \t_0 \right)\biggr]_{s=1}, \label{eqn:G2BC1}\\
\M_2(s=1) &= C_m \t_2 \mid_{s=1} \times \H - \beta l_m^2 \biggl\{ \frac{1}{2} \left(\t_0 \cdot \u_1+\t_1 \cdot \u_0 \right) \notag \\
& \ \ \ \ +\frac{l_m}{3} \left[ \OOmega_1 -(\t_1 \cdot \OOmega_0 +\t_0 \cdot \OOmega_1) \t_0 - (\t_0 \cdot \OOmega_0) \t_1   \right]    \biggr\}_{s=1} \cdot
\end{align}
\normalsize 

From the local balance of force (Eq.~\ref{eqn:G2F}) with the boundary conditions at $s=0$ (Eq.~\ref{eqn:G2BC0}), we find that $\N_2 = (N_{2x}, N_{2y},N_{2z})$ is given by
\small
\begin{align}
N_{2x}(s) &= \frac{\h}{\sqrt{1+\h^2}} \left(\gamma^{-1}-1\right) \times \notag \\ 
& \ \ \ \ \left[ J(s) + \frac{\tilde{U}_1 s}{\sqrt{1+\h^2}} + \frac{\h}{\sqrt{1+\h^2}} H(s)+\frac{\h}{\sqrt{1+\h^2}}C_3 s \right] \notag \\
& \ \ \ \  +H(s)+ C_3 s,\\
N_{2y}(s) &= -b_1 s,\\
N_{2z} (s) &= \frac{\gamma^{-1}-1}{\sqrt{1+\h^2}} \times \notag \\ 
& \ \ \ \ \biggl[ J(s) + \frac{\tilde{U}_1 s}{\sqrt{1+\h^2}} + \frac{\h}{\sqrt{1+\h^2}} H(s)+\frac{\h}{\sqrt{1+\h^2}}C_3 s \biggr] \\ \notag
& \ \ \ \ + U_1 s,
\end{align}
\normalsize 
where $C_3$ is an unknown integration constant and we define the functions
\begin{align}
F(s) &= \frac{d (b_0 \phi_1)}{d s} = \frac{\h s^4}{24 \sqrt{1+\h^2}}-\frac{C_1 s^3}{6} +B_1,\\
G(s) &= \int^s_0 F(s') ds' = \frac{\h s^5}{120 \sqrt{1+\h^2}}-\frac{C_1 s^4}{24}+B_1 s,\\
H(s) &= \int^s_0 G(s') ds' = \frac{\h s^6}{720 \sqrt{1+\h^2}}-\frac{C_1 s^5}{120}+B_1 \frac{s^2}{2},\\
J(s) &= \int^s_0 b_0 (s')  F(s') ds'.
\end{align}
Examining the force components $N_{2x}$ and $N_{2z}$, we have two unknowns, namely $U_1$ and $C_3$. These two unknowns are determined by applying the boundary conditions at $s=1$ (Eq.~\ref{eqn:G2BC1}), yielding a $2\times2$ system of equations
\small
\begin{align}
\begin{pmatrix}
A_{11} & A_{12}\\  
A_{21} & A_{22}
\end{pmatrix}
\begin{pmatrix}
\tilde{U}_1\\  
C_3
\end{pmatrix}
=
\begin{pmatrix}
\h \Phi_1 + \beta l_m G(1) + \frac{\beta l_m^2}{2} F(1)+H(1)\\
\Phi_1
\end{pmatrix},
\end{align}
where
\begin{align}
A_{11} &= A_{22} = -\left[ \frac{h(\gamma^{-1}-1)}{1+\h^2} + \frac{\h \beta l_m (\gamma_m^{-1}-1)}{1+\h^2}\right],\\
A_{12} &= - \left[ \beta l_m + \frac{\h^2\beta l_m (\gamma_m^{-1}-1)}{1+\h^2}+1+\frac{\h^2 (\gamma^{-1}-1)}{1+\h^2}\right],\\
A_{21} &= -\left[  \beta l_m + \frac{\beta l_m (\gamma_m^{-1}-1)}{1+\h^2}+1+\frac{\gamma^{-1}-1}{1+\h^2} \right], \\
\Phi_1 &= \frac{\gamma^{-1}-1}{\sqrt{1+\h^2}} \left[ J(1)+\frac{\h}{\sqrt{1+\h^2}}H(1)\right] \notag \\
& \ \ \ \ + \frac{\beta l_m (\gamma_m^{-1}-1)}{\sqrt{1+\h^2}} \left[ \frac{\h}{\sqrt{1+\h^2}}G(1)+b_0(1) F(1)\right].
\end{align}
\normalsize 

Upon solving this system of linear equation, we arrive at an explicit formula for the leading order swimming speed 
\begin{align}
\tilde{U} &= \tilde{U}_1 \Gamma + O(\Gamma^2) \\ 
& = \frac{\h^2}{1440 \left(1+\h^2\right)^{3/2} (1+\beta l_m  )^2 (\gamma \beta l_m  +\gamma_m)} \times \notag \\
 & \ \ \ \ \biggl\{ 5\gamma \gamma_m [1+  \beta l_m (4+3 l_m) ]^2-4\gamma \beta l_m  \left[ 4+\beta l_m (13+9 l_m) \right] \notag \\
& \ \ \ \  -5 \gamma_m-\gamma_m \beta l_m \left[24+30 l_m + \beta l_m \left(28+84 l_m+45 l_m^2\right)  \right]  \biggr\} \Gamma \notag \\
& \ \ \ + O(\Gamma^2). \label{eqn:lowSpSpeed}
\end{align}

Again, we can verify that when we have isotropic drag $\gamma=\gamma_m=1$, then no swimming is possible, $\tilde{U}_1=0$.

\begin{figure}[t]
\centering
\includegraphics[width=0.4\textwidth]{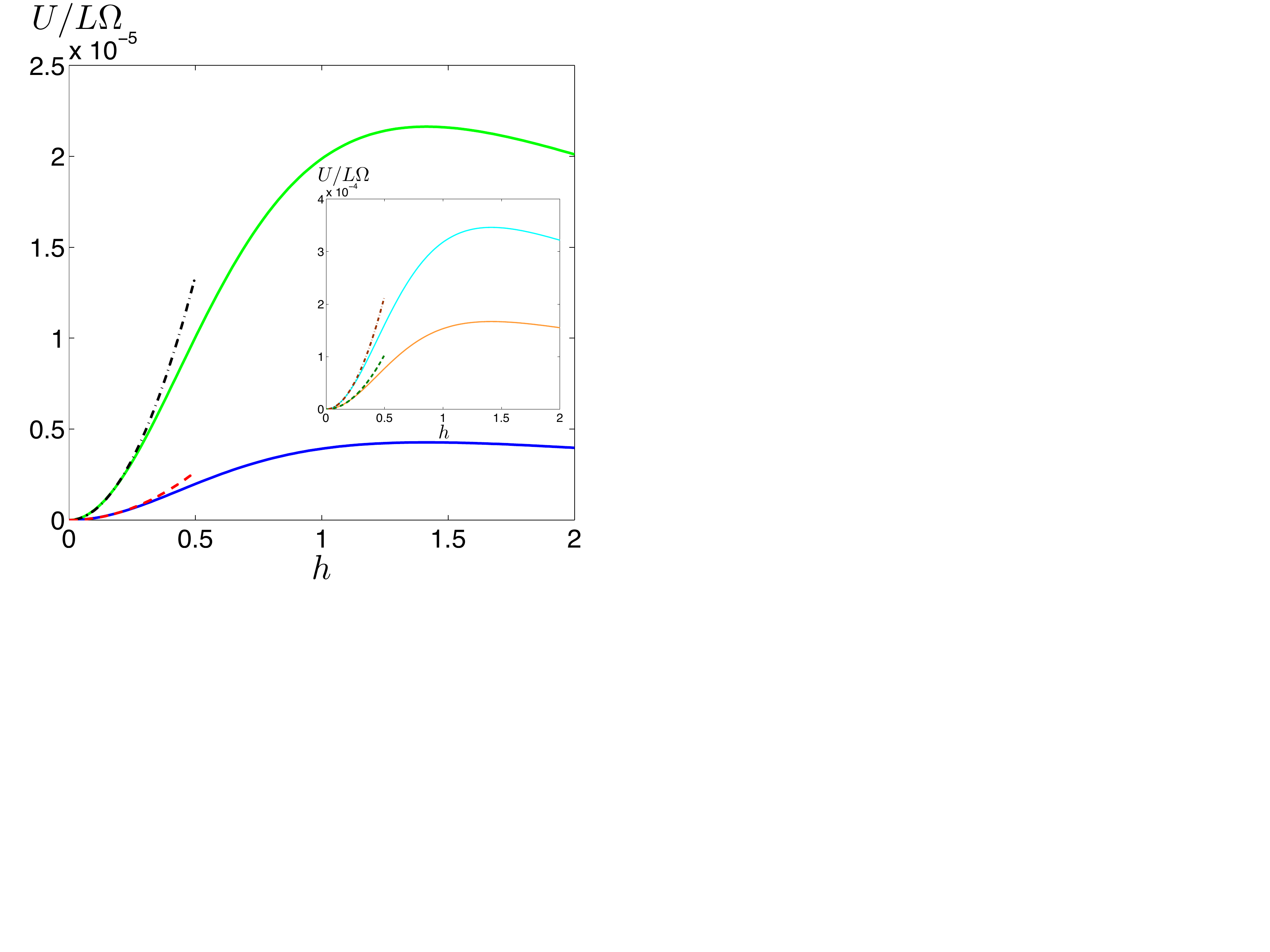}
\caption{\label{fig:VaryH} Variation of the dimensionless swimming speed, $U/L\Omega$, with the relative magnetic field strength, $\h$, for different sperm numbers based on the low-$\Sp$ calculations: $\Sp=0.2$ (dark blue solid line), $\Sp=0.3$ (light green solid line). The red dotted ($\Sp=0.2$) and black dash-dotted ($\Sp=0.3$) lines are the corresponding results from the small-$\h$ calculations. Inset: Same plot for $\Sp=0.5$ (dark orange solid line) and $\Sp=0.6$ (light blue solid line). The green dotted ($\Sp=0.5$) and brown dash-dotted ($\Sp=0.6$) lines are the corresponding results from the small-$\h$ calculations.}

\end{figure}
\subsection{Variation with the relative magnetic field strength, h}
First, one can see that $C_m$, which is the ratio of the characteristic magnetic torque to the characteristic elastic torque, does not enter the formula for $\tilde{U}_1$ (Eq.~\ref{eqn:lowSpSpeed}), meaning that the absolute value of the magnetic field strength or the dipole moment strength has not yet played a role in the swimming speed at low sperm numbers. However, the relative strength of the rotating and constant magnetic field, $\h=H_1/H_0$, has an interesting effect here. From the small-$\h$ asymptotic analysis in the main text, we knew swimming occurs at $O(\h^2)$ (Eq.~\ref{eqn:speed}), and hence the swimming speed (both dimensional or dimensionless) scales quadratically with $\h$, $U \sim \h^2$, for $\h \ll 1$. This is confirmed by examining Eq.~\eqref{eqn:lowSpSpeed}, where we have 
$\tilde{U} \sim {\h^2}/{(1+\h^2)^{3/2}} = \h^2 + O(\h^3)$ when expanded for small $\h$. When the dimensionless swimming speeds from the two asymptotic analyses are plotted against $\h$ for small sperm numbers (Fig.~\ref{fig:VaryH}, different lines represent results at various sperm numbers, see the caption for details), we see an excellent agreement when $\h$ is small, illustrating that the swimming speed does increase quadratically with $\h$ for small $\h$ (the dotted lines are the small-$\h$ asymptotic results). From the low sperm number results (solid lines), the swimming speed then experiences a maximum when $\h$ continues to increase (the small-$\h$ results are no longer valid in this regime), and eventually decreases with further increase in $\h$. From the analytical expression (Eq.~\ref{eqn:lowSpSpeed}), we see that the maximum swimming speed occurs at $\h = \sqrt{2} \approx 1.41$.

\begin{figure}[t]
\centering
\includegraphics[width=0.4\textwidth]{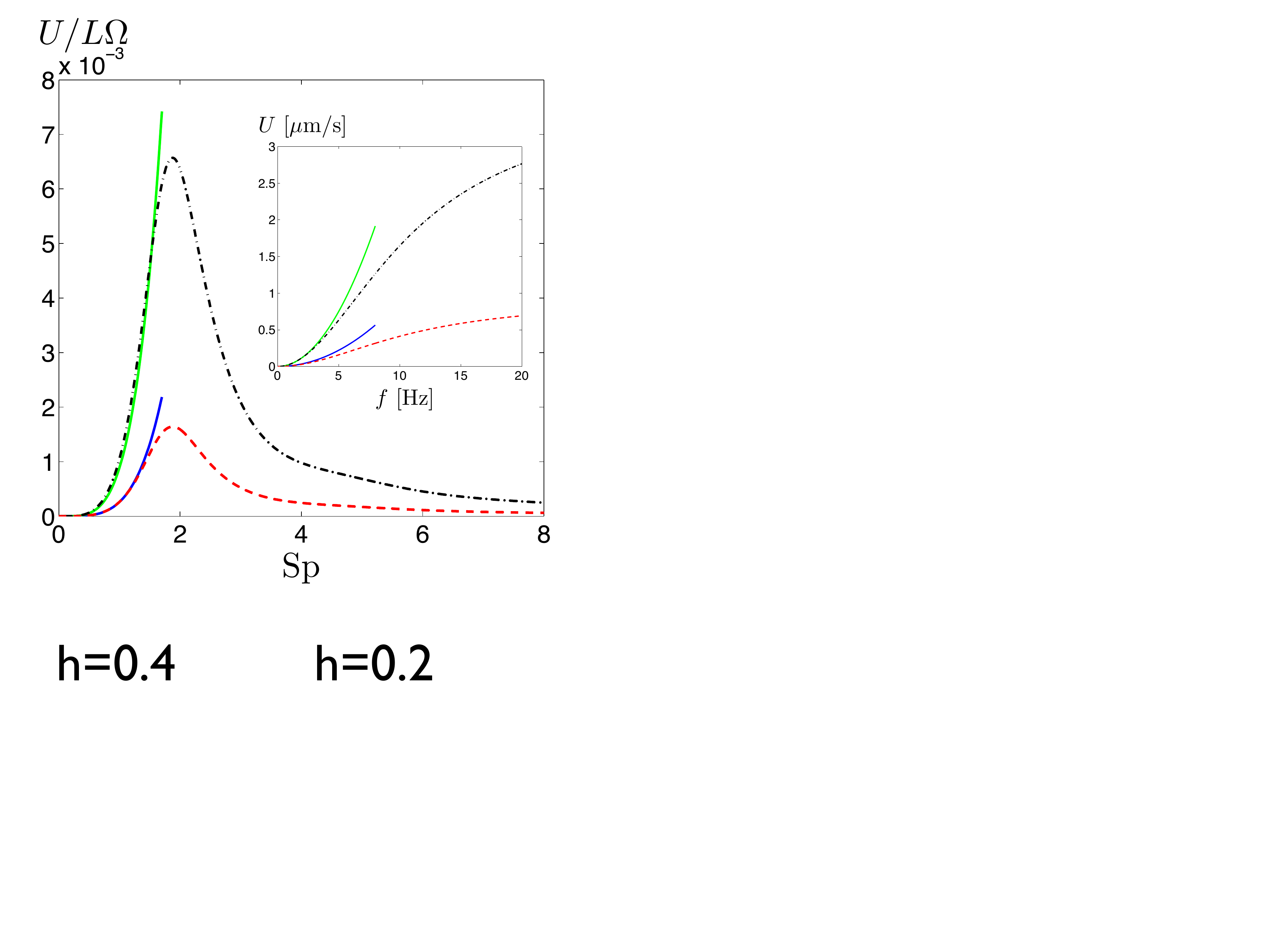}
\caption{\label{fig:VarySpK} Variation of the dimensionless swimming speed, $U/L\Omega$, with the sperm number, $\Sp$, based on the low-$\Sp$ calculations: $\h=0.2$ (dark blue solid line), $\h=0.4$ (light green solid line). The red dotted ($\h=0.2$) and black dash-dotted ($\h=0.4$) lines are the corresponding results from the small-$\h$ calculations. Inset: Variation of the dimensional swimming speed $U$ with the frequency $f$ based on the low-$\Sp$ calculations: $\h=0.2$ (dark blue solid line), $h=0.4$ (light green solid line). The red dotted ($\h=0.2$) and black dash-dotted ($\h=0.4$) lines are the corresponding results from the small-$\h$ calculations. A bending stiffness of $A=3.6 \times 10^{-24}$Nm$^2$ is used in the speed-frequency plot (inset).}
\end{figure}

\subsection{Variation with the sperm number, $\Sp$}
From the small-$\h$ asymptotic analysis in the main text, we have illustrated the dependence of the dimensionless swimming speed on the sperm number, $\Sp$ (Fig.~\ref{fig:SpermNumber}). Here, via the low sperm number asymptotic results (Eq.~\ref{eqn:lowSpSpeed}), we see quantitatively that the dimensionless swimming speed scales as the fourth power of the sperm number, $\tilde{U}=U/L \Omega \sim \Gamma \sim \Sp^4$, for low $\Sp$. Since $\Sp^{4}  \propto f $, this also means that the dimensional swimming speed scales quadratically with the frequency  $U \sim f^2$, for small $f$. We confirm this result by plotting the variation of the dimensionless swimming speed with the sperm number (Fig.~\ref{fig:VarySpK}), and the variation of the dimensional swimming speed with the frequency (Fig.~\ref{fig:VarySpK} inset). We compare the low-Sp asymptotic results ($\h=0.2$, dark blue solid line; $\h=0.4$, light green solid line) with the corresponding small-h asymptotic results $\h=0.2$ (red dotted line) and $\h=0.4$ (black dash-dotted line), so that the small-h asymptotic assumption ($\h \ll 1$) is expected to be valid. We see the results from the two asymptotic analyses agree with each other for sufficiently low sperm numbers (Fig.~\ref{fig:VarySpK}), or frequencies (Fig.~\ref{fig:VarySpK} inset), confirming the scaling $U/L\Omega \sim \Sp^4$ ($U \sim f^2$), at low sperm numbers (low frequencies).

To summarize, in this appendix, with the help of a low-$\Sp$ asymptotic analysis, we have confirmed the scaling $U \sim \h^2$ for small $\h$ in the main text, and established the complementary scalings $U/L\Omega \sim \Sp^4$ for small $\Sp$, or equivalently $U \sim f^2$ for small $f$. Note that the results in this appendix are valid for very small $\Sp$ numbers, and hence are not expected to be useful for experimental comparison. In addition, $C_m$ is assumed to be $O(1)$ throughout the calculations. Since $C_m \propto l_m$, the results here are invalid for $l_m \ll 1$ and $l_m \gg 1$.

\bibliography{nano}

\end{document}